\documentclass[aps,physrev,preprint,groupedaddress]{revtex4-2} 

\usepackage{graphicx}
\usepackage{dcolumn}
\usepackage{bm}

\begin{document}

\title{
On three-cluster resonance structure of hypernuclei \\ $_{\Lambda}^{7}$He,
$_{\Lambda}^{7}$Li and $_{\Lambda}^{7}$Be 
}%
\author{N. Kalzhigitov}
\email{knurto1@gmail.com}
\author{S. Amangeldinova}
\author{V. O. Kurmangaliyeva}
\affiliation{Department of Theoretical and Nuclear Physics, Farabi University, Almaty 050040, Kazakhstan}

\author{V. S. Vasilevsky}
\email{vsvasilevsky@gmail.com}
\affiliation{Bogolyubov Institute for Theoretical Physics,\\ 14-b Metrolohichna str. Kyiv 03143, Ukraine
} 

\date{\today}%

\begin{abstract}
The bound and resonance states of hypernuclei $_{\Lambda}^{7}$He, $_{\Lambda}^{7}%
$Li and $_{\Lambda}^{7}$Be are studied within a three-cluster model. Within
this model, hypernuclei $_{\Lambda}^{7}$He, $_{\Lambda}^{7}$Li and $_{\Lambda
}^{7}$Be are considered as three-cluster systems $^{4}$He+$^2n$+$\Lambda$,
$^{4}$He+$d$+$\Lambda$, $^{4}$He+$^2p$+$\Lambda$, respectively. Special
attention is paid to determining resonance states in the three-cluster continuum
of these hypernuclei and to study their nature. One semi-realistic
nucleon-nucleon potential is employed to determine the internal structure of the
clusters $^{4}$He, $^2n$, $d$ and $^2p$, and their interaction. Three versions of
the nucleon-hyperon potential, known as the YNG potential, are employed to
determine the interaction of the listed clusters with the lambda hyperon. A set of
very narrow and fairly wide resonance states is found in the three-cluster
continuum of  $_{\Lambda}^{7}$He, $_{\Lambda}^{7}$Li and $_{\Lambda}^{7}$Be.
The narrowest resonance states were detected  in $_{\Lambda}^{7}$He and
$_{\Lambda}^{7}$Li and their total width does not exceed 10 keV. The dominant
decay channels of these resonance states are revealed. 
\end{abstract}

\maketitle


\section{Introduction}

In this paper, we study the discrete and continuous spectrum of
seven-particle hypernuclei $_{\Lambda}^{7}$He, $_{\Lambda}^{7}$Li and
$_{\Lambda}^{7}$Be. These hypernuclei will be investigated within a
three-cluster microscopic model.
For the hypernuclei of interest, we selected the following three-cluster
partitions:%
\begin{eqnarray*}
_{\Lambda}^{7}\text{He}  &  =&^{4}\text{He}+^{2}n+\Lambda,\\
_{\Lambda}^{7}\text{Li}  &  =& ^{4}\text{He}+d+\Lambda,\\
_{\Lambda}^{7}\text{Be}  &  = & ^{4}\text{He}+^{2}p+\Lambda.
\end{eqnarray*}
 
Hypernuclei  $_{\Lambda}^{7}$He, $_{\Lambda}^{7}$Li and $_{\Lambda}^{7}$Be
have been a subject of different theoretical methods such as 
many-cluster models \cite{2009PhRvC..80e4321H, 2009PrPNP..63..339H, 2015PhRvC..91e4316H,
 2018PhRvC..97b4330K, 2023PhRvC.107e4302M}, the Gamow shell model
\cite{2025PhLB..86839708L}, ab initio no-core shell models
\cite{2014PhRvL.113s2502W, 2020EPJA...56..301L, 2018PhRvC..97f4315W,
 2026PhLB..87240074K}, mean-field models \cite{1993PhRvC..48..889G,
2001PhRvC..64d4301V}. These investigations were aimed at reproducing existing
experimental data, revealing the nature of the detected bound states, and
predicting new states and their parameters.  A large variety of realistic and
semi-realistic nucleon-nucleon and nucleon-hyperon potentials are involved in
these investigations. Consider some of these publications.

In
Ref. \cite{2019PTEP.2019g3D01K}, the antisymmetrized molecular dynamics model
has been used to study the properties of bound states of a large number of p-shell hypernuclei, including  $_{\Lambda}^{7}$Li.
The spin-orbit components of the hyperon-nucleon interaction were tested to
reproduce the order and energy of the observable states. In Ref.
\cite{2018PhRvC..97b4330K} a microscopic three-cluster model has been applied
to study a large set of p-shell hypernuclei. Special attention has been paid
to the energy shift and size difference of energy levels due to the
interaction of the lambda hyperon with a core nucleus. 

A four-cluster model has been used in \cite{2006PhRvC..74e4312H,
2009PhRvC..80e4321H,  2015PhRvC..91e4316H,
 2023PhRvC.107e4302M}  seven-particle hypernuclei. These nuclei were
considered as structureless $^{4}$He interacting with two nucleons and the
lambda-hyperon. The Pauli principle was treated with the orthogonality
condition model. In Ref. \cite{2009PhRvC..80e4321H}, this model was applied to 
study  hypernuclei $_{\Lambda}^{7}$He, $_{\Lambda}^{7}$Li and $_{\Lambda}^{7}%
$Be. The influence of odd and even components of the central $\Lambda N$
interaction and the symmetric and antisymmetric components of the 
spin-orbit $\Lambda N$ interaction on the relative position of the ground and excited states
was considered in detail. In Ref. \cite{2023PhRvC.107e4302M}, the complex
scaling method was utilized to determine the resonance structure of neutron-rich
He isotopes. A similar model was used in \cite{2015PhRvC..91e4316H} to study the
spectrum of $_{\Lambda}^{7}$He. In this case, the complex scaling method was
used to determine the bound and resonance states  of  $^{6}$He and
$_{\Lambda}^{7}$He. Two very broad resonance states were discovered  in $_{\Lambda
}^{7}$He just above the $^{4}$He+$n+n+\Lambda$ threshold.

The Gamow shell model (GSM) has been applied in Ref.
\cite{2025PhLB..86839708L} to study a list of neutron-rich He
hyper-isotopes from $_{\Lambda}^{6}$He to $_{\Lambda}^{9}$He, among them
$_{\Lambda}^{7}$He. In this model, they were considered as inert $^{4}$He core
interacting with a cloud of neutrons and the lambda hyperon.  Spectra of bound
states were calculated. The proton, neutron, and hyperon root-mean-square radii
were also determined for the ground states of these hypernuclei.

The three-cluster model, formulated in \cite{1983PThPh..70..189M} and applied to the study of several hypernuclei including $_{\Lambda}^{7}$Li, well reproduced the
existing experimental data. This model also used the orthogonality condition
methodology to treat the Pauli principle. The basis of oscillator functions is
used to expand wave functions of two-cluster subsystems ($^{6}$Li=$^{4}$He+d
for $_{\Lambda}^{7}$Li) and the wave function of relative motion of the lambda
hyperon and two-cluster subsystem.

Analysis of publications on properties of light hypernuclei shows that the main attention of different theoretical models has been mainly paid to their bound states, and the resonance states have been rarely investigated. There is also a lack of experimental and theoretical information on resonance states in these hypernuclei, which decay into two or three clusters. We wish to fill this gap by employing a three-cluster microscopic model. Thus, our main aim is to detect three-cluster resonance states in these hypernuclei.

More specifically, in the present work, we study the structure of the bound and resonance
states of the $^{7}_{\Lambda}$He, $_{\Lambda}^{7}$Li and
$_{\Lambda}^{7}$Be hypernuclei. Our prime aim is to investigate resonance
states formed in three-cluster continua of these hypernuclei. To achieve this aim, we
adopted a three-cluster model, which was formulated in Ref.
\cite{2001PhRvC..63c4606V} and designed to study the bound state of light
nuclei and their continuous spectrum states decaying on three fragments (clusters). This method has been successfully applied to study the structure of
light atomic nuclei and especially the resonance states in the three-cluster continuum
of these nuclei \cite{2010PPN....41..716N, 2012PhRvC..85c4318V,
 2017PhRvC..96c4322V, 2018PhRvC..98b4325V}. This method has also been
applied to study properties of resonance states in $^{12}$C and
especially the $0^{+}$ resonance state, 
known as the Hoyle-state. Knowledge of the Hoyle-state properties, accumulated in Ref.
\cite{2012PhRvC..85c4318V}, allowed one to find the Hoyle-analog states in a
set of light nuclei \cite{2018PhRvC..98b4325V}. As in these papers, we refer
to this method as AMHHB, which means the Algebraic version of the resonating
group Method which uses the Hyperspherical Harmonics Basis to numerate channels of a three-cluster system and to implement correct boundary
conditions for decay of a compound nucleus on three clusters.

It is worthwhile noticing that three-cluster resonance states are not exotic, they are observed in a large set of light nuclei. This set is comprised by the
so-called  Borromean nuclei (such as $^{6}$He, $^{9}$Be, $^{11}$Li, $^{12}%
$C), which have a three-cluster decay threshold as the lowest of all possible
two- and three-cluster decay thresholds.  To this set, one can also add nuclei
that have no bound states but have a rather large number of resonance states that reside in a three-cluster continuum of these nuclei. They are, for example,
$^{6}$Be, $^{9}$Be. 
Three-cluster resonance states have been the subject of
numerous experimental and theoretical investigations. In Ref. \cite{2018PhRvC..98b4325V} one can find a list of theoretical methods which
are employed to study the existing three-cluster resonance states
and to predict their appearance in exotic nuclei. 

The structure of the paper is as follows. In Sec. \ref{Sec:Method} provides a brief overview 
of the main ideas of the microscopic model
involved in the calculations. In Sec. \ref{Sec:Results} we fix all input
parameters of our model and discuss spectra of bound states of hypernuclei $_{\Lambda}^{7}$He,
$_{\Lambda}^{7}$Li and $_{\Lambda}^{7}$Be. In this section, the parameters of the resonance states and
their dominant decay channels are analyzed.

\section{Method AMHHB \label{Sec:Method}}

To study the structure of light hypernuclei presented as three-cluster systems, we adapt a microscopic model which combines the Resonating Group Method
and the Hyperspherical Harmonics Method. Details of the model and its
application to the study of bound and continuous spectrum states of light nuclei
can be found in Refs. \cite{2001PhRvC..63c4606V, 2001PhRvC..63c4607V,
2007JPhG...34.1955B, 2010PPN....41..716N, 2012PhRvC..85c4318V}.

We start with the seven-particle system (six nucleons and one lambda hyperon)
described with a microscopic Hamiltonian, and then we reduce it to an effective three-body problem by splitting seven particles into three groups
(clusters). Then we assume that we know wave functions that describe the internal
structure of each cluster with acceptable precision. Based on these
assumptions, the wave function of a hypernucleus $_{\Lambda}^{A}Z$ divided into
three clusters $_{\Lambda}^{A}Z=
=^{A_{1}}Z_{1}+^{A_{2}%
}Z_{2}+\Lambda$ ($A=A_{1}+A_{2}+1$, $ Z=Z_{1}+Z_{2}$) is represented as%
\begin{equation}
\Psi_{E,J}=\sum_{L,S}\widehat{\mathcal{A}}\left\{  \left[  \Phi_{1}\left(
A_{1},S_{1},b\right)  \Phi_{2}\left(  A_{2},S_{2},b\right)  \Phi_{3}\left(
\Lambda,S_{3}\right)  \right]  _{S}\psi_{E,LSJ}\left(  \mathbf{x,y}\right)
\right\}  _{J},\label{eq:N001}%
\end{equation}
where the wave function $\psi_{E,LJ}\left(  \mathbf{x},\mathbf{y}\right)  $
describes the relative motion of the clusters, $\Phi_{1}\left(  A_{1},S_{1},b\right)
$ is the wave function of an alpha particle ($A_{1}=$4$,$ $S_{1}$=0),
$\Phi_{2}\left(  A_{2},S_{2},b\right)  $ is the wave function of a two-nucleon subsystem:
deuteron ($A_{2}$=2, $S_{2}$=1), dineutron or diproton ($A_{2}$=2,  $S_{2}%
$=0). Many-particle wave functions $\Phi_{1}\left(  A_{1},S_{1},b\right)  $
and $\Phi_{2}\left(  A_{2},S_{2},b\right)  $, being the eigenfunctions of the
many-particle shell-model Hamiltonian with harmonic oscillator interaction,
explicitly depend on the oscillator radius (length) $b$. As the lambda hyperon is
considered as a structureless particle, the wave function $\Phi_{3}\left(
\Lambda,S_{3}\right)  $ represents the spin part of the lambda hyperon
function ($A_{3}=m_{\Lambda},$ $S_{2}$=1/2). The antisymmetrization operator
$\widehat{\mathcal{A}}$ permutes only nucleons and thus makes antisymmetric a
wave function of $^{6}$Li ($^{6}$He and $^{6}$Be) which is considered as a
two-cluster system $^{4}$He+$d$ ($^{4}$He+$^2n$ and $^{4}$He+$^2p$). Two vectors
$\mathbf{x}$ and $\mathbf{y}$ denote a possible set of Jacobi vectors.
In this paper, the vector $\mathbf{x}$ determines  the distance between two
selected clusters $A_{1}$ and $A_{2}$, while the vector $\mathbf{y}$
represents the displacement of the lambda hyperon with respect to the center of mass
of two ordinary clusters. Or explicitly%
\begin{eqnarray}
\mathbf{x} &  = &\sqrt{\frac{A_{1}A_{2}}{A_{1}+A_{2}}}\left[  \frac{1}{A_{1}%
}\sum_{i\in A_{1}}\mathbf{r}_{i}-\frac{1}{A_{2}}\sum_{j\in A_{2}}%
\mathbf{r}_{i}\right]  ,\label{eq:M003}\\
\mathbf{y} &  =& \sqrt{\frac{m_{\Lambda}\left(  A_{1}+A_{2}\right)  }%
{m_{\Lambda}+A_{1}+A_{2}}}\left[  \mathbf{r}_{\Lambda}-\frac{1}{A_{1}+A_{2}%
}\sum_{i\in A_{1}+A_{2}}\mathbf{r}_{i}\right]  ,\nonumber
\end{eqnarray}
where $m_{\Lambda}$= 1.188 is the mass of the lambda hyperon in units of nucleon mass.

It is very convenient to use the $LS$ coupling scheme for three interacting
$s$-clusters. In this scheme, the total spin $S$ is a vector sum of individual cluster
spins, and the total orbital momentum $\widehat{\mathbf{L}}$ is
also a vector sum $\widehat{\mathbf{L}}=\widehat{\mathbf{l}}_{x}%
+\widehat{\mathbf{l}}_{y}$ of the partial orbital momenta $\widehat
{\mathbf{l}}_{x}$ and $\widehat{\mathbf{l}}_{y}$, associated with the Jacobi
vectors $\mathbf{x}$ and $\mathbf{y}$, respectively. The total angular
momentum $J$ is a vector sum of the total orbital momentum $L$ and the total
spin $S$.

To simplify obtaining wave functions of discrete and continuous spectrum
states and scattering parameters and to implement correct boundary conditions,
we transit from the Jacobi vectors $\mathbf{x}$ and $\mathbf{y}$ to the
hyperspherical coordinates, which consist of hyperradius $\rho$ and five
hyperspherical angles which we denote as $\Omega_{5}$. The hyperradius $\rho$
is defined in a standard way
\begin{equation}
\rho=\sqrt{\mathbf{x}^{2}+\mathbf{y}^{2}}. \label{eq:N002A}%
\end{equation}
The set of hyperspherical angles $\Omega_{5}$ is composed of five angular
variables: the hyperspherical angle $\theta$ that determines the relative lengths
of the Jacobi vectors $\mathbf{x}$ and $\mathbf{y}$
\begin{equation}
x=\rho\cos\theta,\quad y=\rho\sin\theta, \label{eq:N002B}%
\end{equation}
two angles $\theta_{x}$ and $\phi_{x}$, determining the orientation of the vector
$\mathbf{x}$, and two other angles $\theta_{y}$ and $\phi_{y}$, determining the
orientation of the vector $\mathbf{y}$ in the space. Note that the angles
$\left\{  \theta_{x},\phi_{x}\right\}  $ describe the rotation of a two-cluster
subsystem and the angles $\left\{  \theta_{y},\phi_{y}\right\}  $ describe the
rotation of the third cluster around the center of mass of the two-cluster
subsystem. Five hyperspherical angles are capable of describing any shape and any
orientation (or rotation) of a triangle connecting the centers of mass of three
clusters, and the hyperradius $\rho$ determines any size of that triangle.

With the hyperspherical coordinates, we can represent the three-cluster wave
function (\ref{eq:N001}) in the following form
\begin{equation}
\Psi_{E,J}=\sum_{c}\widehat{\mathcal{A}}\left\{  \left[  \Phi_{1}\left(
A_{1},S_{1},b\right)  \Phi_{2}\left(  A_{2},S_{2},b\right)  \Phi_{3}\left(
\Lambda,S_{3}\right)  \right]  _{S}\psi_{E,c}\left(  \rho\right)
\mathcal{Y}_{c}\left(  \Omega_{5}\right)  \right\}  _{J}, \label{eq:N003}%
\end{equation}
where $\mathcal{Y}_{c}\left(  \Omega\right)  $ stands for the product
\begin{equation}
\mathcal{Y}_{c}\left(  \Omega\right)  =\chi_{K}^{\left(  \lambda,l\right)
}\left(  \theta\right)  \left\{  Y_{\lambda}\left(  \widehat{\mathbf{x}%
}\right)  Y_{l}\left(  \widehat{\mathbf{y}}\right)  \right\}  _{LM_{L}}
\label{eq:M012}%
\end{equation}
and represents a hyperspherical harmonic for a three-cluster channel, the quantum number $c$ is a
multiple index $c=\left\{  K;\lambda,l;L,S\right\}  $ classifying channels of
the three-cluster system and involving the hypermomentum $K$, partial orbital
momenta $\lambda$ and $l$ associated with  Jacobi vectors $\mathbf{x}$ and
$\mathbf{y}$, respectively, and the total orbital momentum $L$. The explicit
form of the function $\chi_{K}^{\left(  \lambda,l\right)  }\left(
\theta\right)  $ is
\begin{equation}
\chi_{K}^{\left(  \lambda,l\right)  }\left(  \theta\right)  =N_{K}^{\left(
\lambda,l\right)  }\cos^{\lambda}\theta\sin^{l}\theta~P_{n}^{\left(
l+1/2,\lambda+1/2\right)  }, \label{eq:M013}%
\end{equation}
where $P_{n}^{\left(  \alpha,\beta\right)  }\left(  x\right)  $ is the Jacobi
polynomial, and
\begin{eqnarray*}
N_{K}^{\left(  \lambda,l\right)  }  &  =& \sqrt{\frac{2\left(  K+2\right)
n!\left(  n+l+\lambda+1\right)  !}{\Gamma\left(  n+l+3/2\right)  \Gamma\left(
n+\lambda+3/2\right)  }},\\
n  &  =& \left(  K-l-\lambda\right)  /2.
\end{eqnarray*}
The hyperspherical harmonics $\mathcal{Y}_{c}\left(  \Omega_{5}\right)  $ form
a complete set of functions on a five-dimensional sphere and thus account for all
kinds of motion of a three-cluster system. The components of the many-channel
hyperradial wave function $\left\{  \psi_{E,c}\left(  \rho\right)  \right\}  $
have to be determined by solving the Schr\"{o}dinger equation with the
selected nucleon-nucleon nucleon-hyperon potentials.

\subsection{Three-cluster equation}

For three structureless particles, one obtains the infinite set of differential
equations%
\begin{equation}
\sum_{\widetilde{c}}\left[  \delta_{c,\widetilde{c}}\widehat{T}_{K}%
+V_{c,\widetilde{c}}\left(  \rho\right)  \right]  \psi_{E,\widetilde{c}%
}\left(  \rho\right)  =E\psi_{E,c}\left(  \rho\right)  , \label{eq:N008}%
\end{equation}
where%
\begin{equation}
\widehat{T}_{K}=-\frac{\hbar^{2}}{2m}\left[  \frac{\partial^{2}}{\partial
\rho^{2}}+\frac{5}{\rho}\frac{\partial}{\partial\rho}-\frac{K\left(
K+4\right)  }{\rho^{2}}\right]  . \label{eq:N008A}%
\end{equation}
The matrix $\left\Vert V_{c,\widetilde{c}}\left(  \rho\right)  \right\Vert $ of the effective potential energy is determined as  the matrix elements of the interaction
$\widehat{V}$ between the hyperspherical harmonics%
\begin{equation}
V_{c,\widetilde{c}}\left(  \rho\right)  =\left\langle \mathcal{Y}%
_{c}\left\vert \widehat{V}\right\vert \mathcal{Y}_{\widetilde{c}}\right\rangle
, \label{eq:N009}%
\end{equation}
where integration is performed over all hyperspherical angles $\Omega_{5}$.
The operator $\widehat{V}$ includes the sum of the nucleon-nucleon and
nucleon-hyperon potentials
\[
\widehat{V}=\sum_{i<j\in A_{1}+A_{2}}\widehat{V}_{NN}\left(  i,j\right)
+\sum_{i\in A_{1}+A_{2}}\widehat{V}_{N\Lambda}\left(  i,\Lambda\right)
\]
If both clusters $A_{1}$ and $A_{2}$ have electric charges (Z$_{1}\neq$0,
Z$_{2}\neq$0), then we have the additional contribution
\begin{equation}
V_{c,\widetilde{c}}^{\left(  C\right)  }\left(  \rho\right)  =\left\langle
\mathcal{Y}_{c}\left\vert \sum_{i<j\in A_{1}+A_{2}}\widehat{V}_{C}\left(
i,j\right)  \right\vert \mathcal{Y}_{\widetilde{c}}\right\rangle
=\frac{Z_{c,\widetilde{c}}e^{2}}{\rho} \label{eq:N009A}%
\end{equation}
from the Coulomb interaction of protons%
\[
\widehat{V}_{C}\left(  i,j\right)  =\frac{1}{4}\left(  1+\widehat{\tau}%
_{iz}\right)  \left(  1+\widehat{\tau}_{jz}\right)  \frac{e^{2}}{\left\vert
\mathbf{r}_{i}-\mathbf{r}_{j}\right\vert }.
\]
to the potential energy $V_{c,\widetilde{c}}\left(  \rho\right)  $
(\ref{eq:N009}). The quantity $Z_{c,\widetilde{c}}$ can be called the
effective charge. Assuming that at large values of hyperradius the effective
potential $V_{c,\widetilde{c}}\left(  \rho\right)  $, which originated from a short
range particle-particle interaction, is negligibly small and omitting
non-diagonal elements of the effective charge (by putting $Z_{c,\widetilde{c}%
}=0$ for $c\neq\widetilde{c}$), we obtain an asymptotic part of the channel
Hamiltonian%
\begin{equation}
\widehat{H}_{c}^{\left(  A\right)  }=\left\{  -\frac{\hbar^{2}}{2m}\left[
\frac{\partial^{2}}{\partial\rho^{2}}+\frac{5}{\rho}\frac{\partial}%
{\partial\rho}-\frac{K\left(  K+4\right)  }{\rho^{2}}\right]  +\frac
{Z_{c,c}e^{2}}{\rho}\right\}  . \label{eq:N011}%
\end{equation}
The eigenfunctions of this Hamiltonian describing incoming $\psi_{c}^{\left(
-\right)  }\left(  \rho,\eta_{c}\right)  $ and outgoing $ \psi_{c}^{\left(
+\right)  }\left(  \rho,\eta_{c}\right)  $ hyperradial waves can be easily
found and expressed through the Whittaker functions (see chapter 13.1 in Ref.
\cite{kn:abra})
\begin{equation}
\psi_{c}^{\left(  \pm\right)  }\left(  \rho,\eta_{c}\right)  =\sqrt{\frac{\pi
}{2}}\frac{1}{\rho^{5/2}}W_{\mp i\eta_{c},K+2}\left(  \mp2ik\rho\right)  ,
\label{eq:N012}%
\end{equation}
where%
\[
k=\sqrt{\frac{2mE}{\hbar^{2}}}%
\]
and $\eta_{c}$ is the Sommerfeld parameter for the three-cluster system%
\[
\eta_{c}=\frac{m}{\hbar^{2}}\frac{Z_{c,c}e^{2}}{k}.
\]
Thus, the boundary conditions or the asymptotic form of many-channel wave functions can be expressed in the form%
\[
\psi_{E,c}\left(  \rho\right)  =\delta_{c_{0},c}\psi_{c}^{\left(  -\right)
}\left(  \rho,\eta_{c}\right)  -S_{c_{0},c}\psi_{c}^{\left(  +\right)
}\left(  \rho,\eta_{c}\right)  ,
\]
where $c_{0}$ stands for an incoming channel, $S_{c_{0},c}$ is an element of
the scattering $S$-matrix.

When the internal structure of clusters of three-cluster systems and the Pauli principle are taken into account, we obtain the set of integro-differential equations:%
\begin{eqnarray}
&  \sum_{\widetilde{c}}\left[  \delta_{c,\widetilde{c}}\widehat{T}_{K}%
\psi_{E,\widetilde{c}}\left(  \rho\right)  +\int d\widetilde{\rho}%
\widetilde{\rho}^{5}V_{c,\widetilde{c}}\left(  \rho,\widetilde{\rho}\right)
\psi_{E,\widetilde{c}}\left(  \widetilde{\rho}\right)  \right] \label{eq:N020}%
\\
&  =E\sum_{\widetilde{c}}\int d\widetilde{\rho}\widetilde{\rho}^{5}%
N_{c,\widetilde{c}}\left(  \rho,\widetilde{\rho}\right)  \psi_{E,\widetilde
{c}}\left(  \widetilde{\rho}\right)  ,\nonumber
\end{eqnarray}
which involves nonlocal potentials $V_{c,\widetilde{c}}\left(  \rho
,\widetilde{\rho}\right)  $ and the norm kernel $N_{c,\widetilde{c}}\left(
\rho,\widetilde{\rho}\right)  $. This system of equations can be obtained from
the many-particle Schr\"{o}dinger equations with the help of the projection
operator%
\begin{equation}
\widehat{P}_{c}\left(  \rho\right)  =\widehat{\mathcal{A}}\left\{  \left[
\Phi_{1}\left(  A_{1}\right)  \Phi_{2}\left(  A_{2}\right)  \Phi_{3}\left(
A_{3}\right)  \right]  _{S}\delta\left(  \rho-\overline{\rho}\right)
\mathcal{Y}_{c}\left(  \Omega_{5}\right)  .\right\}  \label{eq:N021}%
\end{equation}
Applying this operator to the unit operator, we obtain the norm kernel
$N_{c,\widetilde{c}}\left(  \rho,\widetilde{\rho}\right)  $
\begin{equation}
N_{c,\widetilde{c}}\left(  \rho,\widetilde{\rho}\right)  =\left\langle
\widehat{P}_{c}\left(  \rho\right)  |\widehat{P}_{\widetilde{c}}\left(
\widetilde{\rho}\right)  \right\rangle . \label{eq:N022}%
\end{equation}
In Eq. (\ref{eq:N022}), integration is performed over all spatial coordinates
(the Jacobi vectors) and over all spin and isospin coordinates as well. The
matrix of potential energy is related to the matrix elements of the
microscopic Hamiltonian $\widehat{H}$ by the relation
\begin{equation}
V_{c,\widetilde{c}}\left(  \rho,\widetilde{\rho}\right)  =\left\langle
\widehat{P}_{c}\left(  \rho\right)  \left\vert \widehat{H}\right\vert
\widehat{P}_{\widetilde{c}}\left(  \widetilde{\rho}\right)  \right\rangle
-\delta_{c,\widetilde{c}}\widehat{T}_{K}\delta\left(  \rho-\widetilde{\rho
}\right)  . \label{eq:N023}%
\end{equation}
The system of Eq. (\ref{eq:N020}) can be directly solved by reducing it to a reasonable finite number of  three-cluster channels $N_{c}$ and with
the boundary conditions determined above. The solutions of the system give us the definite set of matrix elements of the $S$ matrix. They describe all kinds
of elastic and inelastic processes in a three-cluster system.

Within the present model, a wave function (\ref{eq:N001}) of a three-cluster system is expanded over an infinite set of cluster oscillator functions
$\left\vert n_{\rho},c\right\rangle $%
\[
\Psi_{E,LJ}=\sum_{n_{\rho},c}C_{n_{\rho},c}^{E,J}\left\vert n_{\rho
},c\right\rangle ,
\]
where%
\begin{eqnarray}
&  \left\vert n_{\rho},c\right\rangle =\left\vert n_{\rho},K;\lambda
,l;L\right\rangle \label{eq:N010}\\
&  =\widehat{\mathcal{A}}\left\{  \Phi_{1}\left(  A_{1}\right)  \Phi
_{2}\left(  A_{2}\right)  \Phi_{3}\left(  A_{3}\right)  R_{n_{\rho}K}\left(
\rho,b\right)  \mathcal{Y}_{c}\left(  \Omega_{5}\right)  \right\}  ,\nonumber
\end{eqnarray}
$\mathcal{Y}_{c}\left(  \Omega_{5}\right)  $ is a familiar hyperspherical
harmonic with the quantum numbers $c=\left\{  K,l_{x},l_{y},L\right\}  $ and
$R_{n_{\rho},K}\left(  \rho,b\right)  $ is an oscillator function%
\begin{eqnarray}
R_{n_{\rho},K}\left(  \rho,b\right)   &  = &\left(  -1\right)  ^{n_{\rho}%
}\mathcal{N}_{n_{\rho},K} x^{K}\exp\left\{  -\frac{1}{2}x^{2}\right\}  L_{n_{\rho}}
^{K+3}\left(  x^{2}\right) ,  \label{eq:N010A}\\
x  &  =& \rho/b,\quad\mathcal{N}_{n_{\rho},K}=b^{-3}\sqrt{\frac{2\Gamma\left(
n_{\rho}+1\right)  }{\Gamma\left(  n_{\rho}+K+3\right)  }},\nonumber
\end{eqnarray}
and $b$ is an oscillator length.

In this case, a set of the integro-differential equations is reduced to a set of the algebraic (matrix) equations%

\begin{equation}
\sum_{\widetilde{n}_{\rho},\widetilde{c}}\left[  \left\langle n_{\rho
},c\left\vert \widehat{H}\right\vert \widetilde{n}_{\rho},\widetilde
{c}\right\rangle -E\left\langle n_{\rho},c|\widetilde{n}_{\rho},\widetilde
{c}\right\rangle \right]  C_{\widetilde{n}_{\rho},\widetilde{c}}^{E,J}=0,
\label{eq:N030}%
\end{equation}
which can be solved more easily by numerical methods than the set of
equations (\ref{eq:N020}). For continuous spectrum states, one has to impose
proper boundary conditions for expansion coefficients $\left\{  C_{n_{\rho}%
,c}^{E,J}\right\}  $. These conditions have been discussed in Ref.
\cite{2001PhRvC..63c4606V, 2018PhRvC..97f4605V} where relations between
the discrete $\left\{  C_{n_{\rho},c}^{E,J}\right\}  $ and continuous
$\left\{  \psi_{E,c}\left(  \rho\right)  \right\}  $ wave functions were
established. By including the asymptotic form of the expansion coefficients
$\left\{  C_{n_{\rho},c}^{E,J}\right\}  $, which is valid for large values of
hyperradial excitations $n_{\rho}\gg1$, we obtain in a closed form the system
of equations that determine both wave functions of a continuous spectrum and the
corresponding $S$ matrix.

The system of equations (\ref{eq:N030}) can be solved numerically by imposing
restrictions on the number of hyperradial excitations $n_{\rho}$ and on the
number of hyperspherical channels $c_{1}$, $c_{2}$, \ldots, $c_{N_{ch}}$. The
diagonalization procedure is used to determine the energies and wave functions of
the bound states. However,  proper boundary conditions have to be
implemented to calculate elements of the scattering $S$-matrix and
corresponding functions of the continuous spectrum. Boundary conditions for wave
functions of democratic and nondemocratic decay of a compound three-cluster
system are thoroughly discussed in Refs \cite{2001PhRvC..63c4606V,
 2018PhRvC..97f4605V}.

Having solved the system of the dynamic equations (\ref{eq:N030}), we obtain the
$N_{ch}\times N_{ch}$ scattering S-matrix $\left\Vert S_{cc^{\prime}%
}\right\Vert $ for the $N_{ch}$ channel system and the corresponding $N_{ch}$ wave
functions. The element $S_{c_{,}c^{\prime}}$ of the $S$-matrix describes the
transition from the initial channel $c$ to the final channel $c^{\prime}$. To analyze processes in the three-cluster continuum, we prefer to use two different
representations of the $S$-matrix. In the first representation, we determine the
phase shift $\delta_{c_{,}c^{\prime}}$ and  inelastic parameter
$\eta_{c_{,}c^{\prime}}$%
\begin{equation}
S_{c_{,}c^{\prime}}=\eta_{c_{,}c^{\prime}}\exp\left\{  2i\delta_{c_{,}%
c^{\prime}}\right\}  . \label{eq:S005}%
\end{equation}
This is a traditional representation for the S-matrix of the many-channel systems. By reducing
the S-matrix $\left\Vert S_{c_{,}c^{\prime}}\right\Vert  $ to diagonal form, we obtain the second representation, which involves the uncoupled eigenchannels, each of
them is determined by the eigenphase shift $\delta_{\alpha}$ or the
$S_{\alpha}$-matrix%
\begin{equation}
S_{\alpha}=\exp\left\{  2i\delta_{\alpha}\right\}  , \label{eq:S006}%
\end{equation}
where $\alpha$ (=1,2,\ldots, $N_{ch}$) numerates the eigenchannels. The relation
between the original $\left\Vert S_{c_{,}c^{\prime}}\right\Vert $ and diagonal
$\left\Vert S_{\alpha}\right\Vert $ forms of the $S$- matrix is
\begin{equation}
S_{c_{,}c^{\prime}}=\sum_{\alpha}U_{\alpha}^{c}S_{\alpha}U_{\alpha}%
^{c^{\prime}} \label{eq:S007}%
\end{equation}
where $\left\Vert U_{\alpha}^{c}\right\Vert $ is an orthogonal matrix. The energy and width of a resonance state are determined from the eigenphase
shifts:%
\begin{equation}
\left.  \frac{d^{2}\delta_{\alpha}}{dE^{2}}\right\vert _{E=E_{R}}%
=0,\qquad\Gamma_{\alpha}=2\left[  \left.  \frac{d\delta_{\alpha}}%
{dE}\right\vert _{E=E_{R}}\right]  ^{-1}. \label{eq:S008}%
\end{equation}
The partial widths $\Gamma_{c\alpha}$, determining the decay of a resonance
state in a specific channel $c$, can be determined in the following way:%
\begin{equation}
\Gamma_{c\alpha}=\left\vert U_{\alpha}^{c}\right\vert ^{2}\Gamma_{\alpha}.
\label{eq:S009}%
\end{equation}
In Ref. \cite{2007JPhG...34.1955B}, one can find  the justifications of such a
determination of the total and partial widths for many-channel systems.

\subsection{Definition of basic quantities}

Having obtained the expansion coefficients for any state of the three-cluster
continuum, we can easily construct its wave function in the coordinate space.
It can be done first of all for the total hyperradial wave function
\begin{equation}
\psi_{E,c}\left(  \rho\right)  =\sum_{n_{\rho}}C_{n_{\rho},c}^{E,J}R_{n_{\rho
},K}\left(  \rho,b\right)  . \label{eq:N033}%
\end{equation}
It can also be  done for the wave function%
\begin{equation}
\psi_{E,LJ}\left(  \mathbf{x},\mathbf{y}\right)  =\sum_{n_{\rho},c}C_{n_{\rho
},c}^{E,J}R_{n_{\rho},K}\left(  \rho,b\right)  \mathcal{Y}_{c}\left(
\Omega_{5}\right)  . \label{eq:N034}%
\end{equation}

To get more information about the state under consideration, we will study
different quantities which can be obtained with the wave function in the discrete (oscillator) or continuous (coordinate) spaces. With wave functions in the discrete oscillator 
 representation, we can determine the weight $W_{sh}$ of the oscillator
functions that belong to the oscillator shell $N_{sh}$ in this wave function:
\begin{equation}
W_{sh}\left(  N_{sh}\right)  =\sum_{n_{\rho},c\in N_{sh}}\left\vert
C_{n_{\rho},c}^{E,J}\right\vert ^{2}. \label{eq:N035}%
\end{equation}
where the summation is performed over all hyperspherical harmonics and
hyperradial excitations obeying the following condition:
\[
N_{os}=2n_{\rho}+K.
\]
Here, $N_{os}$ is fixed. 
It
is convenient to numerate the oscillator shells by $N_{sh}$ ( = 0, 1, 2,
\ldots), which we determine as
\[
N_{os}=2n_{\rho}+K=2N_{sh}+K_{\min},
\]
where $K_{\min}=L$ for the normal parity states $\pi=\left(  -1\right)  ^{L}$ and
$K_{\min}=L+1$ for the abnormal parity states $\pi=\left(  -1\right)  ^{L+1}$.
Thus, we count oscillator shells starting from a "vacuum" shell ($N_{sh}$ =
0) with minimal value of hypermomentum $K_{\min}$ compatible with a given
total orbital momentum $L$.

 We will calculate the weights $W_{sh}$ for both the bound and resonance states.
For a bound state, the wave function is normalized by the condition%
\begin{equation}
\left\langle \Psi_{E,J}|\Psi_{E,J}\right\rangle =\sum_{n_{\rho},c}\left\vert
C_{n_{\rho},c}^{E,J}\right\vert ^{2}=1, \label{eq:N036A}%
\end{equation}
and this quantity $W_{sh}$ determines the probability. For the continuous
spectrum state, when the wave function is normalized by the condition%
\begin{equation}
\left\langle \Psi_{E,J}|\Psi_{\widetilde{E},J}\right\rangle =\sum_{n_{\rho}%
,c}C_{n_{\rho},c}^{E,J}C_{n_{\rho},c}^{\widetilde{E},J}=\delta\left(
k-\widetilde{k}\right)  , \label{eq:N036B}%
\end{equation}
this quantity has a different meaning. It determines the relative contribution
of the different oscillator shells and  also the shape of the resonance wave
function in the oscillator representation.

It is worth noting that oscillator functions have some important
features. The oscillator functions belonging to an oscillator shell $N_{sh}$ allow
one to describe a many-particle system in a finite range of hyperradius
$0<\rho\leq b\sqrt{4N_{sh}+K_{\min}+6}$. Outside this region, these oscillator
functions give a negligible small contribution to the many-particle wave function.
This statement is, for example, demonstrated in Ref. \cite{2018PhRvC..97f4605V}.
Thus, oscillator functions with a small value of $N_{sh}$ describe very
compact configurations of a three-cluster system with all clusters being close
to each other. When $N_{sh}$ is large, the oscillator functions represent 
dispersed (dilute) configurations. There are two principal regimes in these
configurations. The first regime is associated with a two-body type of
asymptotics when two clusters are at a small distance and the third cluster is moved far away. The second regime accounts for the case when all three
clusters are well separated. Taking these into account, we will deduce from an
analysis of shell weights $W_{sh}$ whether a wave function of a bound or
resonance state describes a compact or dispersed three-cluster configuration.

It is necessary to add another important remark. There is a strict
correspondence between the wave function of two-body and two-cluster systems in the oscillator and coordinate spaces, as established in Ref.
\cite{kn:Fil_Okhr, kn:Fil81}.
In Ref. \cite{2001PhRvC..63c4606V},  a correspondence was established between
the expansion coefficients $\left\{  C_{n_{\rho},c}^{E,J}\right\}  $ and the
coordinate wave function $\psi_{E,c}\left(  \rho\right)  $ for three-cluster
systems. It reads as
\begin{equation}
C_{n_{\rho},c}^{E,J}\approx\sqrt{2}\rho_{n}^{2}\psi_{E,c}\left(  \rho
_{n}\right)  , \label{eq:N040}%
\end{equation}
where
\[
\rho_{n}=b\sqrt{4n_{\rho}+2K+6}%
\]
is a coordinate of the classical turning point in the six-dimensional harmonic
oscillator. Formally, this correspondence is valid for  large values of
$n_{\rho}$, however, in real cases this correspondence is also valid for
moderate values of $n_{\rho}$. From this correspondence one may deduce that
the weights $W_{sh}\left(  N_{sh}\right)  $ for large values of $N_{sh}$ are
proportional to the squared modulus of the wave function $\Psi_{E,J}$ (Eq.
(\ref{eq:N001})) at a discrete value of hyperradius $\rho=b$ $\sqrt
{4N_{sh}+2K_{\min}+6}$. On the other hand, the correspondence (\ref{eq:N040})
confirms the above-mentioned statement that the weights $W_{sh}\left(
N_{sh}\right)  $ with a small value of $N_{sh}$ describe a compact
configuration of a three-cluster system, while the weights $W_{sh}\left(
N_{sh}\right)  $ for large values of $N_{sh}$ determine the probability
for a dispersed configuration of a three-cluster system.

By employing the wave function in the coordinate space, we determine the
correlation function
\begin{equation}
D\left(  x,y\right)  =x^{2}y^{2}\int\left\vert \psi_{E,LJ}\left(
\mathbf{x},\mathbf{y}\right)  \right\vert ^{2}d\widehat{\mathbf{x}}%
d\widehat{\mathbf{y}} \label{eq:N037}%
\end{equation}
and average distances $R_{\Lambda}$ and $R_{B}$ between clusters
\begin{eqnarray}
R_{\Lambda}  &  = & \sqrt{\frac{\left(  A_{1}+A_{2}+m_{\Lambda}\right)  }{\left(
A_{1}+A_{2}\right)  m_{\Lambda}}}\sqrt{\int y^{2}\left\vert \psi_{E,LJ}\left(
\mathbf{x},\mathbf{y}\right)  \right\vert ^{2}d\mathbf{x}d\mathbf{y}%
},\label{eq:N038A}\\
R_{B}  &  =& \sqrt{\frac{\left(  A_{1}+A_{2}\right)  }{A_{1}A_{2}}}\sqrt{\int
x^{2}\left\vert \psi_{E,LJ}\left(  \mathbf{x},\mathbf{y}\right)  \right\vert
^{2}d\mathbf{x}d\mathbf{y}}.\label{eq:N038B} 
\end{eqnarray}
With these notations, $R_{B}$ determines an average distance between the
alpha-particle and the deuteron (dineutron or diproton), while $R_{\Lambda}$
determines an average distance between the lambda hyperon and the center of mass
of a two-cluster subsystem. Note that in Eq. (\ref{eq:N037}) integration is
performed over unit vectors $\widehat{\mathbf{x}}\ $ and $\widehat{\mathbf{y}}$, while in Eqs. (\ref{eq:N038A}) and (\ref{eq:N038B}) integration is carried
out over all Jacobi vectors or all hyperspherical coordinates.

Note that the correlation function $D\left(  x,y\right)  $ can be determined
both for the bound and resonance states. However, the average distances
$R_{\Lambda}$ and $R_{B}$ can be calculated for the bound states only, since the
 integrals in Eqs. (\ref{eq:N038A}) and (\ref{eq:N038B})
diverge for the resonance states. In Ref. \cite{2023UkrJPh..68..3K} we suggested to extend the definition of average distances $R_{\Lambda}$ and $R_{B}$ to the resonance
states. For
this aim, we restricted the integration within the internal part of the
resonance wave functions, which were normalized to unity. Recall that the
internal part of a wave function is represented in the region (0$\leq\rho
\leq\rho_{\max}$ in the coordinate space or 0$\leq n_{\rho}\leq N^{\left(
i\right)  }$ in the oscillator space), where distances between clusters are
relatively small and the effects of intercluster interactions are very strong. This
 definition of $R_{\Lambda}$ and $R_{B}$ allows us to study the shape of
the triangle, composed of three interacting clusters, but not its size. By
comparing the average distances $R_{\Lambda}$ and $R_{B}$ for different resonances
of the same or another (hyper)nucleus, we obtain more information on the structure of
the resonance wave functions.

\section{Results\label{Sec:Results}}

There are a few input parameters that we have to fix to carry out calculations of
the compound hypernuclei. The key ingredients of the present calculations are
nucleon-nucleon and nucleon-lambda potentials. To study the spectra of hypernuclei
$_{\Lambda}^{7}$He, $_{\Lambda}^{7}$Li and $_{\Lambda}^{7}$Be we employ the
modified Hasegawa-Nagata (HNP) \cite{potMHN1, potMHN2} as the nucleon-nucleon (NN)
interaction and the so-called YNG potential \cite{1994PThPS.117..361Y}
(\cite{2006PhRvC..74e4312H}) as the nucleon-hyperon ($N\Lambda$) interaction.
Note that the YNG potential has three different versions, which are denoted as
NF, ND and NS versions (potentials). In Ref. \cite{2021NuPhA101622325N} the
NF version of this potential has been used to study the bound states of the
hypernucleus $^{9}_{\Lambda}$Be, and its resonance states formed in the two-cluster continuum.
These studies have been performed within a slightly different three-cluster model, which involves
boundary conditions for the decay of a compound system onto two clusters. This model
with the YNG-NF potential reasonably reproduced the spectrum of bound states and
shed some light on the structure of bound and resonance states obtained. This
encouraged us to use this potential to study the structure of $_{\Lambda}^{7}%
$He, $_{\Lambda}^{7}$Li and $_{\Lambda}^{7}$Be. Additionally, we decided to
use two other versions of the YNG potential, in order to investigate the influence
of the shape of a nucleon-hyperon ($N\Lambda$) interaction on the discrete and
continuous spectrum of these hypernuclei.

After selecting NN and $N\Lambda$ potentials, the oscillator length (radius)
$b$  has to be fixed, which is the only free parameter of our model. We selected
it to minimize the threshold energy of the three-cluster channel. Within such
types of calculations (not only in the present model), there are also a few
parameters associated with nucleon-nucleon and nucleon-hyperon interactions,
variations of which are used to reproduce some important features of a
compound system or its two-cluster subsystems. The Majorana exchange parameter $m$ and the
intensity of the spin-orbit interaction $f_{LS}$ of the Hasegawa-Nagata NN
potential are very often used to adjust some selected properties of a compound
system or two-cluster subsystems. The selected YNG potentials depend on the
parameter $k_{F}$ - the Fermi momentum, which is also used as a variation
parameter. In the present calculation, the parameters $m$ and $f_{LS}$ are
selected to optimize or make  a more realistic description of the subsystems,
which consist of two ordinary clusters: $^{4} $He+$^2n$ in $_{\Lambda}^{7}$He
and $^{4}$He+$d$ in $_{\Lambda}^{7}$Li. The Fermi momentum $k_{F}$ is selected
to reproduce the ground state $_{\Lambda}^{7}$He and $_{\Lambda}^{7}$Li. For
the hypernucleus $_{\Lambda}^{7}$Be we use the same values of $m$, $f_{LS}%
$ and $k_{F}$ as for $_{\Lambda}^{7}$He. This will allow us to study the
influence of the Coulomb interaction on the bound and resonance states in
$_{\Lambda}^{7}$Be compared to $_{\Lambda}^{7}$He. Such a choice of the
parameters $m$, $f_{LS}$ and $k_{F}$ leads to the same nuclear and
hypernuclear interaction of three clusters in both hypernuclei, and the only
difference between them originates from the Coulomb interaction.

To correctly describe the  states of the three-cluster continuum, we employ all
hyperspherical harmonics with the hypermomentum $K$ that runs from $K_{\min} $
to $K_{\max}$, where $K_{\max}$=14 for the positive parity states, and
$K_{\max}$=13 for the negative parity states. Minimal value of the
hypermomentum $K_{\min}=L$ for the normal parity states $\pi=\left(
-1\right)  ^{L}$, and $K_{\min}=L+1$ for the abnormal parity states
$\pi=\left(  -1\right)  ^{L+1}$.

It is important to note that the sum of the partial orbital momenta
$l_{x}+l_{y}$  also runs  from $K_{\min}$ to $K_{\max}$, however, the
difference $\left\vert l_{x}-l_{y}\right\vert $ should be less than or equal to the
total orbital momentum $L$. It follows from this that the partial orbital
momentum $l_{x}$ associated with the rotation of two-cluster subsystems
$^{4}$He+$^2n$ ,$^{4}$He+$d$ and $^{4}$He+$^2p$ can have not only zero value and
thus account not only for their ground states, but also their excited 
states. Thus, the present model effectively takes  into account the
polarizability of all binary subsystems. 

Throughout the present paper, the energies of bound and resonance states in the
hypernuclei of interest are determined from the corresponding three-cluster threshold.

\subsection{Optimal parameters}

In Table \ref{Tab:OptimParA7} we collect  input parameters which are
associated with nuclear forces: the oscillator length $b$ is chosen to
minimize the energy of the three-cluster threshold $ ^{4}$He$+^2n+\Lambda$ and $^{4}$He$+d+\Lambda$ channels; a small adjustment $\Delta m$ of the
Majorana parameters of the HNP is used to reproduce the energy of the ground state
 in $^{6}$He and $^{6}$Li subsystems, the intensity of the spin-orbit
interaction $f_{LS}$ is fixed to reproduce position of the very narrow 3$^{+}
$ resonance state in $^{6}$Li. For the subsystems $^{6}$He and $^{6}$Be we
use the original intensity of the spin-orbit interaction, i.e. $f_{LS}$=1. With
these input parameters, we determine the spectrum of bound and resonance states of
$^{6}$He and $^{6}$Li, which is obtained in a two-cluster approximation and is
displayed in Table \ref{Tab:KeyBinSystems}. In this table, we also display the
root-mean-square mass (rms) radii $R_{m}$ and average distances $R_{B}$ between the
clusters. The latter is determined by using an algorithm, suggested in Ref.
\cite{2023UkrJPh..68..3K}, and will be used later to analyze the results of
compound hypernuclei. The energy of the 1$^{+}$ bound states and the energies of the
resonance states in $^{6}$Li are well reproduced with the selected input
parameters. The widths of the 3$^{+}$, 2$^{+}$ and 1$^{+}$ resonance states are also
in good agreement with the experimental data. For $^{6}$He, the results are not so
impressive, especially for the 2$^{+}$ resonance state, which actually lies
in the three-cluster continuum. This discrepancy for the 2$^{+}$ resonance states is
also obtained in a three-cluster model \cite{2015UkrJPh..60.201P} with other
potentials. We do not display the same quantities calculated for $^{6}$Be in the
two-cluster approximation, as this nucleus has no bound state. In the
two-cluster model used, both the $0^{+}$ "ground state" and the first excited
2$^{+}$ state of $^{6}$Be are obtained as pseudo-bound states.%

\begin{table}[b]
\caption{\label{Tab:OptimParA7}
The key input parameters of the present model for hypernuclei
$^{7}_{\Lambda}$He, $^{7}_{\Lambda}$Li, $^{7}_{\Lambda}$Be.}%
\begin{ruledtabular}
\begin{tabular}{ccccc}
$_{\Lambda}^{A}Z$ & 3C & $b$, fm & $\Delta m$ & $f_{LS}$\\\hline
$_{\Lambda}^{7}$He & $^{4}\text{He}+^{2}n+\Lambda$ & 1.399 & 0.0579 &
1.000\\
$_{\Lambda}^{7}$Li & $^{4}\text{He}+d+\Lambda$ & 1.357 & -0.0013 &
0.348\\
$_{\Lambda}^{7}$Be & $^{4}\text{He}+^{2}p+\Lambda$ & 1.399 & 0.0579 &
1.000\\
\end{tabular}
\end{ruledtabular}
\end{table}%
%

\begin{table}[b]
\caption{\label{Tab:KeyBinSystems}
Main parameters of two-cluster subsystems $^6$He and $^6$Li
determined with the selected input parameters and compared with available
experimental data. Energy $E$ and width $\Gamma$ are given in MeV and mass rms radius is in fm.}%
\begin{ruledtabular}
\begin{tabular}{ccccccccc}
$^{A}Z$ &  & \multicolumn{4}{c}{2C model} & \multicolumn{3}{c}{Exp.}%
\\
& $J^{\pi}$ & $E$ & $\Gamma$ & $R_{m}$ & $R_{B}$ & $E$ & $\Gamma$ & $R_{m}$ \\\hline
$^{6}$Li & 1$^{+}$ & -1.474 & - & 2.36 & 4.88 & -1.474 & - & 2.32\\
& 3$^{+}$ & 0.708 & 0.016 & 3.22 & 6.73 & 0.712 & 0.024 & \\
& 2$^{+}$ & 3.018 & 0.991 & 8.88 & 18.82 & 2.838 & 1.30 & \\
& 1$^{+}$ & 4.057 & 2.321 & 9.85 & 20.88 & 4.176 & 1.5 & \\
$^{6}$He & 0$^{+}$ & -0.973 & - & 2.71 & 5.65 & -0.973 & - & 2.48\\
& 2$^{+}$ & 2.144 & 0.457 & 7.86 & 16.66 & 0.824 & 0.113 $\pm$ 20 & \\
\end{tabular}
\end{ruledtabular}
\end{table}%

\subsection{Bound states}

In this section, we consider the spectra of bound states in $_{\Lambda}^{7}$He, $_{\Lambda
}^{7}$Li and $_{\Lambda}^{7}$Be and their properties. Before calculating
bound states, we have to fix the last parameter of the model which
determines the shape and intensity of the nucleon-hyperon interaction. This
parameter is the Fermi momentum $k_{F}$ and is selected to reproduce the
energy of the ground state. As we are using three versions of the YNG potential, we
thus select the Fermi momentum $k_{F}$ for each version and separately for
$_{\Lambda}^{7}$He and $_{\Lambda}^{7}$Li. For hypernucleus $_{\Lambda}^{7}%
$Be, we use the same value of $k_{F}$ as for $^{7}$He. Optimal values of the Fermi
momentum are indicated in Table \ref{Tab:OptimKF}. Despite that the intensity of
the YNG potentials is a polynomial of the second order of the parameter
$k_{F}$, the energies of bound states are linear functions of the parameters. They
behave similarly to the bound states of the lightest hypernuclei, which were
considered in Ref. \cite{2025arXiv250901932K} within a two-cluster model.%

\begin{table}[b]
\caption{\label{Tab:OptimKF}
Optimal values of the Fermi momentum determined for
$^7_{\Lambda}$He and  $^7_{\Lambda}$Li with three versions of the YNG
potential.}%
\begin{ruledtabular}
\begin{tabular}{cccc}
Potential & NF & ND & NS\\\hline
$_{\Lambda}^{7}$He & 0.9470 & 0.9530 & 0.9410\\
$_{\Lambda}^{7}$Li & 0.9465 & 0.9460 & 1.0440\\
\end{tabular}
\end{ruledtabular}
\end{table}%

In Tables \ref{Tab:Spectr7HHe3PotO}, \ref{Tab:Spectr7HBe3PotO} and
\ref{Tab:Spectr7HLi3Pot} we show the spectrum of bound states in hypernuclei
$_{\Lambda}^{7}$He, $_{\Lambda}^{7}$Be and $_{\Lambda}^{7}$Li, respectively.
We also show the mass rms radius $R_{m}$ for each bound state. Analysis of the results, presented in Table \ref{Tab:Spectr7HHe3PotO},  indicates that the 1/2$^{+}$, 3/2$^{+}$ and 5/2$^{+}$ bound states in $_{\Lambda
}^{7}$He are quite compact states, as $R_{m}<$
2.4 fm, while the weakly bound 1/2$^{+}$, 3/2$^{-}$ and 1/2$^{-}$ states are
very dispersed states with large values of $R_{m}>$
4.2 fm. The Coulomb interaction slightly decreases the energy of the bound states in  $_{\Lambda}^{7}$Be (see \ref{Tab:Spectr7HBe3PotO}) and makes them larger.

\begin{table}[b]
\caption{\label{Tab:Spectr7HHe3PotO}
Spectrum of bound states in $^7_{\Lambda}$He, determined with three
YNG potentials. Energy is given in MeV and mass rms radius is in fm.}%
\begin{ruledtabular}
\begin{tabular}{ccccccccc}
\multicolumn{3}{c}{NF, $k_{F}$=0.9470} & \multicolumn{3}{c}{ND, $k_{F}%
$=0.9530} & \multicolumn{3}{c}{NS, $k_{F}$=0.9410}\\
$J^{\pi}$ & $E$ & $R_{m}$ & $J^{\pi}$ & $E$ & $R_{m}$ & $J^{\pi}$ & $E$ &
$R_{m}$\\\hline
1/2$^{+}$ & -6.039 & 2.35 & 1/2$^{+}$ & -6.039 & 2.33 & 1/2$^{+}$ & -6.031 &
2.33\\
3/2$^{+}$ & -3.338 & 2.35 & 3/2$^{+}$ & -3.038 & 2.35 & 3/2$^{+}$ & -2.959 &
2.23\\
5/2$^{+}$ & -2.729 & 2.38 & 5/2$^{+}$ & -3.172 & 2.25 & 5/2$^{+}$ & -2.959 &
2.30\\
1/2$^{+}$ & -1.656 & 4.63 & 1/2$^{+}$ & -1.591 & 4.70 & 1/2$^{+}$ & -1.565 &
4.60\\
\end{tabular}
\end{ruledtabular}
\end{table}%
%

\begin{table}[b]
\caption{\label{Tab:Spectr7HBe3PotO}%
Bound state spectrum of the hypernucleus $^7_{\Lambda}$Be, determined with
three $N\Lambda$ potentials. Energy is given in MeV and mass rms radius is in fm.}%
\begin{ruledtabular}
\begin{tabular}{ccccccccc}
\multicolumn{3}{c}{NF, $k_{F}$=0.9470} & \multicolumn{3}{c}{ND, $k_{F}%
$=0.9530} & \multicolumn{3}{c}{NS, $k_{F}$= 0.9410}\\
$J^{\pi}$ & $E$ & $R_{m}$ & $J^{\pi}$ & $E$ & $R_{m}$ & $J^{\pi}$ & $E$ &
$R_{m}$ \\\hline
1/2$^{+}$ & -4.312 & 2.44 & 1/2$^{+}$ & -4.295 & 2.42 & 1/2$^{+}$ & -4.296 &
2.43\\
3/2$^{+}$ & -1.567 & 2.56 & 5/2$^{+}$ & -1.389 & 2.50 & 5/2$^{+}$ & -1.219 &
2.61\\
5/2$^{+}$ & -1.049 & 2.85 & 3/2$^{+}$ & -1.011 & 2.80 & 3/2$^{+}$ & -1.219 &
2.61\\
1/2$^{+}$ & -0.937 & 4.75 & 1/2$^{+}$ & -0.889 & 4.82 & 1/2$^{+}$ & -0.837 &
4.72\\
\end{tabular}
\end{ruledtabular}
\end{table}%

Despite the Coulomb interaction in $_{\Lambda}^{7}$Li, which is approximately two times smaller than in $_{\Lambda}^{7}$Be, the energies of the $_{\Lambda}^{7}$Li bound states, presented in \ref{Tab:Spectr7HLi3Pot},  are smaller than the corresponding energies of the bound states in   $_{\Lambda}^{7}$He. This results in some reduction of the mass rms radii in $_{\Lambda}^{7}$Li with respect to $_{\Lambda}^{7}$He and $_{\Lambda}^{7}$Be. Indeed, the mass rms radius of the ground state in $_{\Lambda}^{7}$Li is equal  to $R_{m}$=2.2 fm, while $R_{m}$=2.3 fm for the ground state in 
$_{\Lambda}^{7}$He.

\begin{table}[ht]
\caption{\label{Tab:Spectr7HLi3Pot}
Spectrum of bound states in $^7_{\Lambda}$Li. Energy is given in MeV and mass rms is in fm.}%
\begin{ruledtabular}
\begin{tabular}{ccccccccc}
\multicolumn{3}{c}{NF, $k_{F}$=0.9465} & \multicolumn{3}{c}{ND, $k_{F}%
$=0.9460} & \multicolumn{3}{c}{NS, $k_{F}$=1.0440}\\ 
$J^{\pi}$ & $E$ & $R_{m}$ & $J^{\pi}$ & $E$ & $R_{m}$ &
$J^{\pi}$ & $E$ & $R_{m}$ \\\hline
1/2$^{+}$ & -7.099 & 2.21 & 1/2$^{+}$ & -7.097 & 2.20 & 1/2$^{+}$ & -7.089 &
2.18\\
3/2$^{+}$ & -6.959 & 2.21 & 3/2$^{+}$ & -7.150 & 2.19 & 3/2$^{+}$ & -5.476 &
2.27\\
5/2$^{+}$ & -4.844 & 2.06 & 5/2$^{+}$ & -5.007 & 2.04 & 5/2$^{+}$ & -4.696 &
2.05\\
1/2$^{+}$ & -1.461 & 4.53 & 1/2$^{+}$ & -1.248 & 4.33 & 1/2$^{+}$ & -0.800 &
4.50\\
\\
\end{tabular}
\end{ruledtabular}
\end{table}%

\begin{figure}[ht]
\begin{center}
\includegraphics[width=0.75\textwidth]{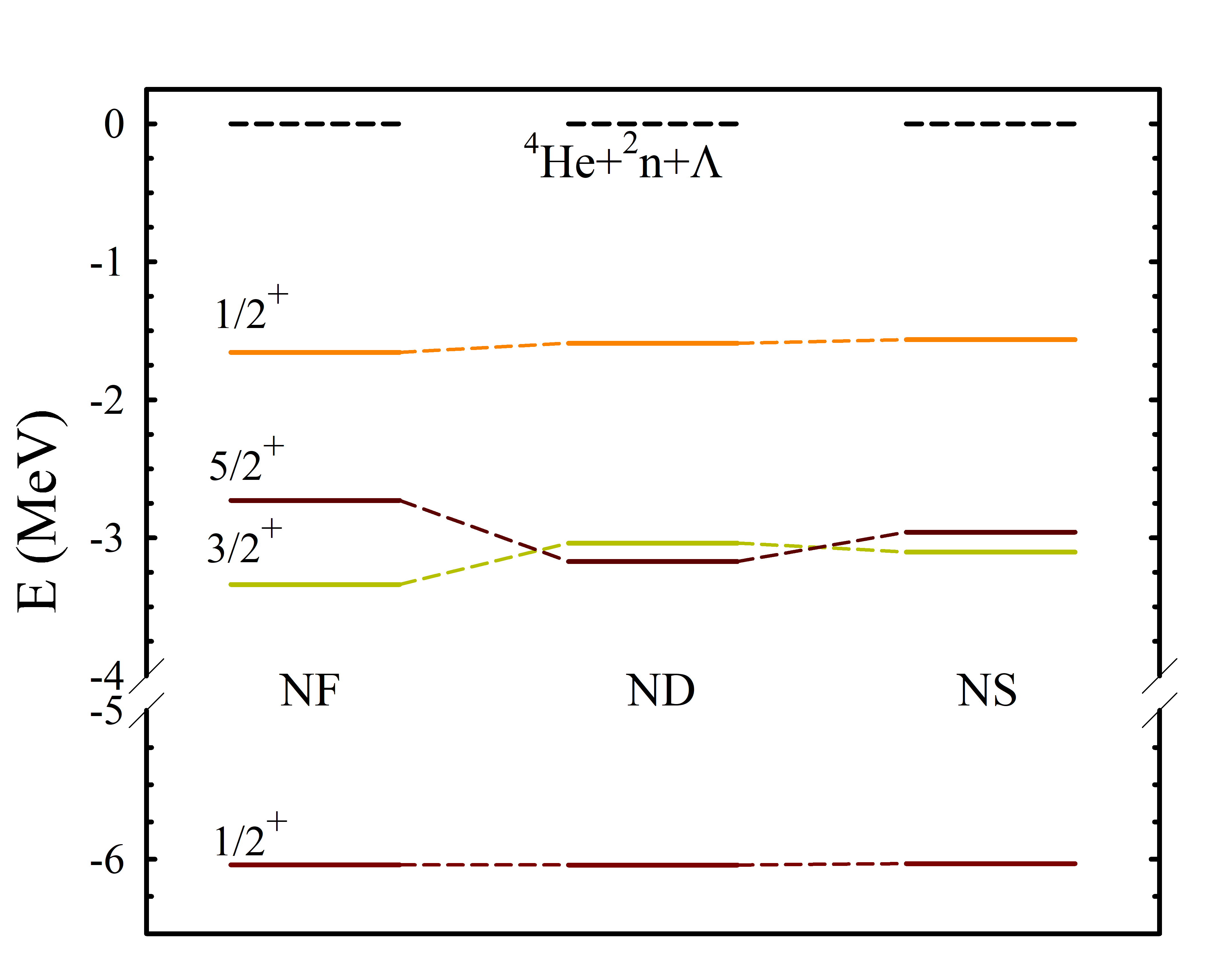}%
\caption{\label{Fig:Figure1}%
Spectrum of bound states in $_{\Lambda}^{7}$He determined with the NF, ND and NS versions of the YNG potential.}%
\end{center}
\end{figure}

\begin{figure}[ht]
\begin{center}
\includegraphics[width=0.75\textwidth]{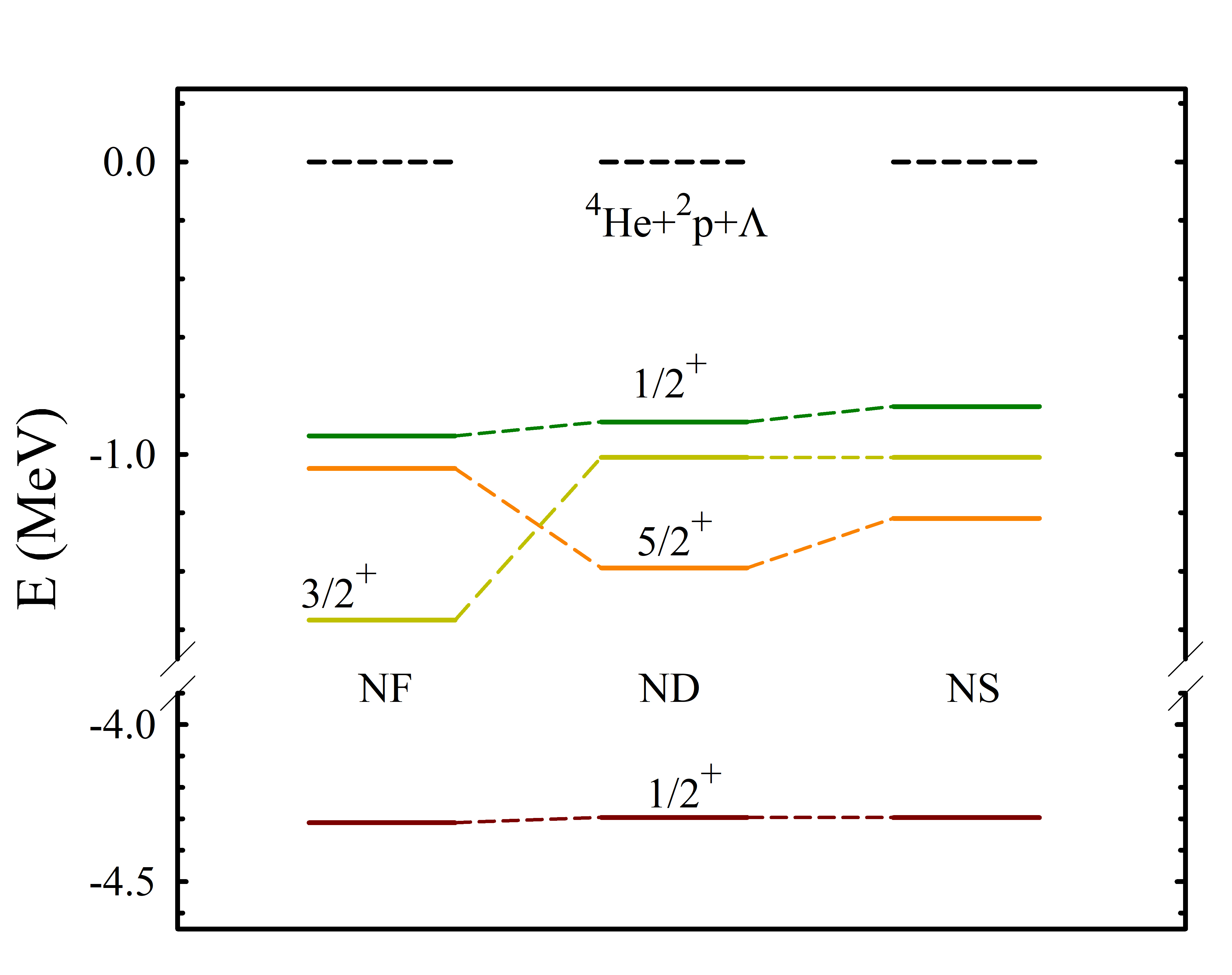}%
\caption{\label{Fig:Figure2}%
Spectrum of bound states in $_{\Lambda}^{7}$Be determined with three versions of the YNG potential.}%
\end{center}
\end{figure}

In Figs. \ref{Fig:Figure1}, \ref{Fig:Figure2} we show the
 relative position of the bound states in hypernuclei $_{\Lambda}^{7}$He and
$_{\Lambda}^{7}$Be. These figures demonstrate how the shape of the YNG potential affects the relative position and order of the bound states. One can see that the position and order of the 5/2$^+$ and 3/2$^+$ states are quite similar for the NF and ND versions and slightly differ for the NS version of the YNG potential.   %

\begin{figure}[ht]
\begin{center}
\includegraphics[width=0.75\textwidth]{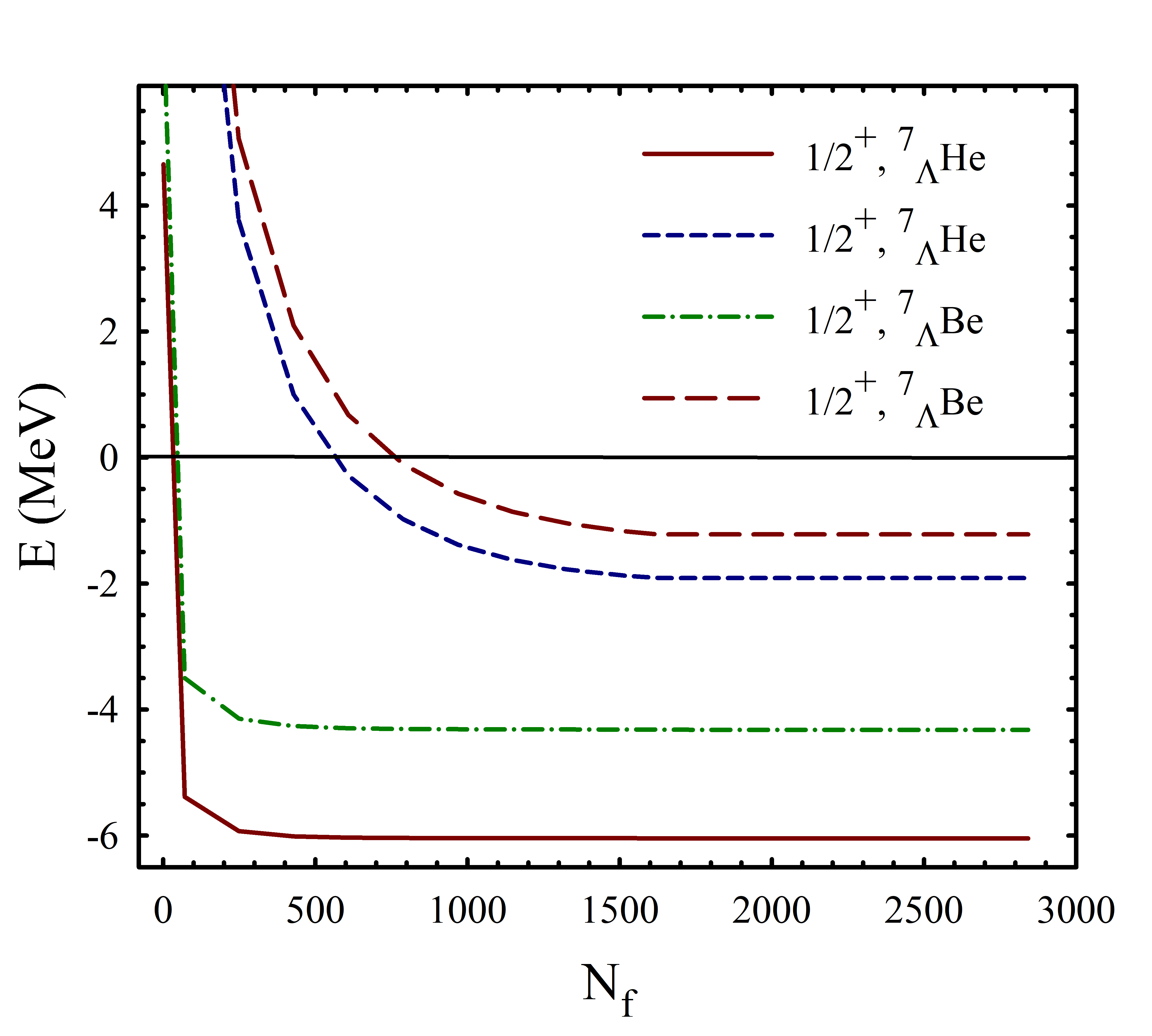}%
\caption{\label{Fig:Figure3}%
Energies of the first and second 1/2$^{+}$ bound states in $_{\Lambda}^{7}$He and $_{\Lambda}^{7}$Be as a function of the number of oscillator functions involved in calculations.}%
\end{center}
\end{figure}

The question may be raised whether we use large bases of hyperradial and hyperspherical functions, which allows us to reach convergence for deeply and weakly bound states. The answer to
this question is demonstrated in Fig. \ref{Fig:Figure3} where the
energies of the ground and first excited 1/2$^{+}$  states of $_{\Lambda}^{7}$He and
$_{\Lambda}^{7}$Be  are depicted as a function of the number of oscillator
functions $N_{f}$. Note  that the oscillator functions with the numbers 1$\leq$
$N_{f}\leq$1617 represent the 1/2$^{+}$ state with the total orbital
momentum $L$=0 and the total spin $S$=1/2, while oscillator functions with the
numbers 1618$\leq N_{F}\leq$2898 represent the total orbital momentum $L$=1 and
total spin $S$=1/2. One can see that a small number of the basis functions
(approximately 500 functions) are needed to provide convergence of the ground
1/2$^{+}$ states in $_{\Lambda}^{7}$He and $_{\Lambda}^{7}$Be, and more than
1500 functions are required to reach convergence for the first excited states
in these hypernuclei. Thus, we use a quite large basis of oscillator functions
which guarantees the convergence of energies even for weakly bound states. Fig.
\ref{Fig:Figure3} demonstrates also  that the ground and first excited
1/2$^{+}$ state in $_{\Lambda}^{7}$He and $_{\Lambda}^{7}$Be are formed by the
total orbital momentum $L$=0 and the orbital momentum $L$=1 plays a negligibly
small role.%

\begin{figure}[ht]
\begin{center}
\includegraphics[width=0.75\textwidth]{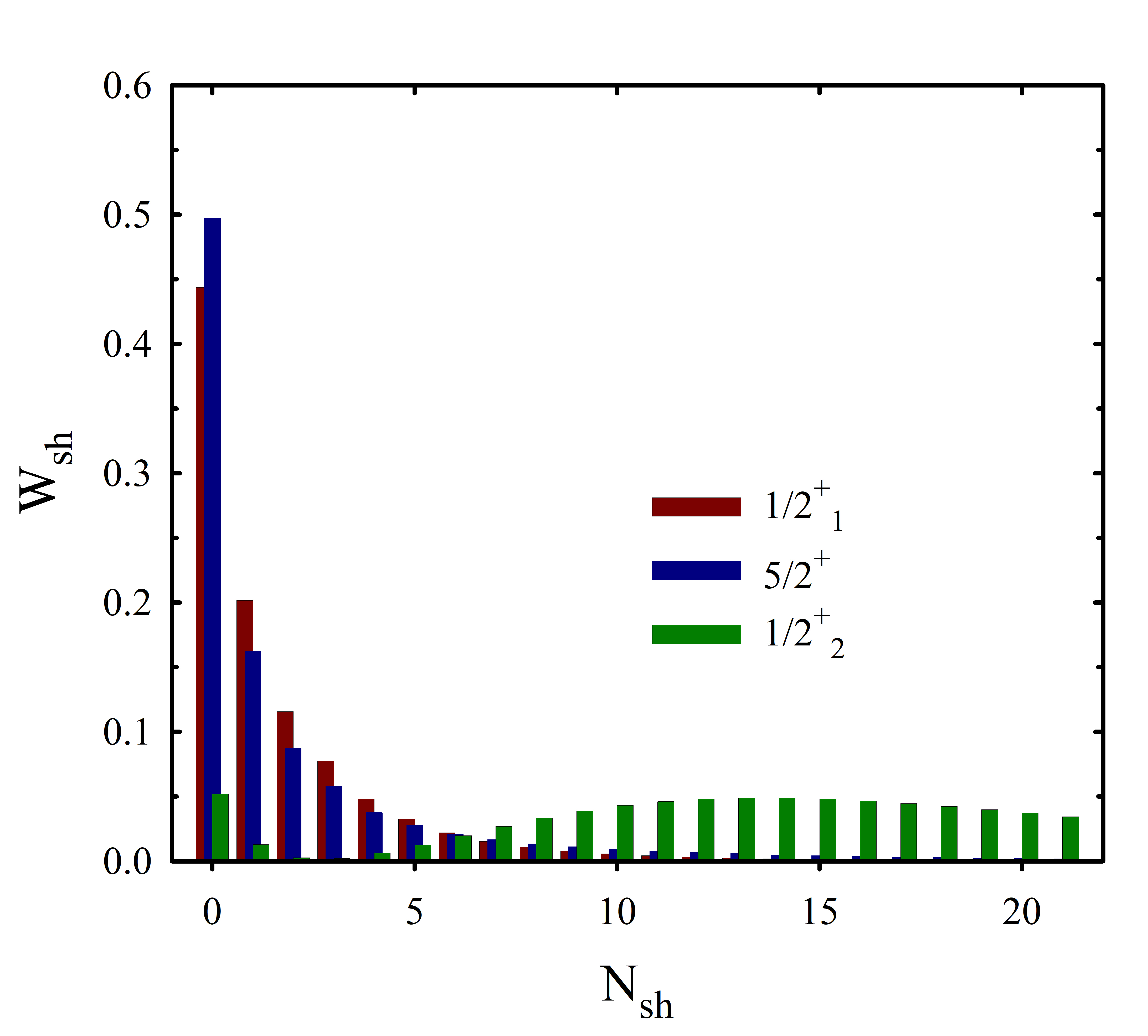}%
\caption{\label{Fig:Figure4} Oscillator shell decomposition of wave functions of bound states in $_{\Lambda}^{7}$He determined with the YNG-NF potential.}%
\end{center}
\end{figure}

Let us analyze the structure of wave functions of bound states. As 
noted above, we analyze the weights of different oscillator shells
$W_{sh}$ in the wave functions of the bound states. In Fig.
\ref{Fig:Figure4} we show the contribution of different shells to the wave
functions of the bound states in $_{\Lambda}^{7}$He, which are determined with the
YNG-NF potential. The lowest oscillator shell $N_{sh}$=0 has the largest
contribution to the wave functions of relatively deeply bound states 1/2$_{1}^{+}
$ and 5/2$^{+}$. This shell is totally associated with the position of the lambda
hyperon with respect to $^{6}$He. The large weight of the oscillator shell
$N_{sh}$=0 indicates that the lambda hyperon with high probability can be found
inside $^{6}$He. The distribution of weights over the oscillator shells shows that
the ground 1/2$_{1}^{+}$ and excited 5/2$^{+}$ are fairly compact states, as
the main contribution comes from the oscillator shells with $N_{sh}\leq$5. The
first 1/2$_{2}^{+}$ excited state, which has a small binding energy, is a very
dispersed state that is formed by the oscillator shells with a large value
$N_{sh}>$5.%

\begin{figure}[ht]
\begin{center}
\includegraphics[width=0.75\textwidth]{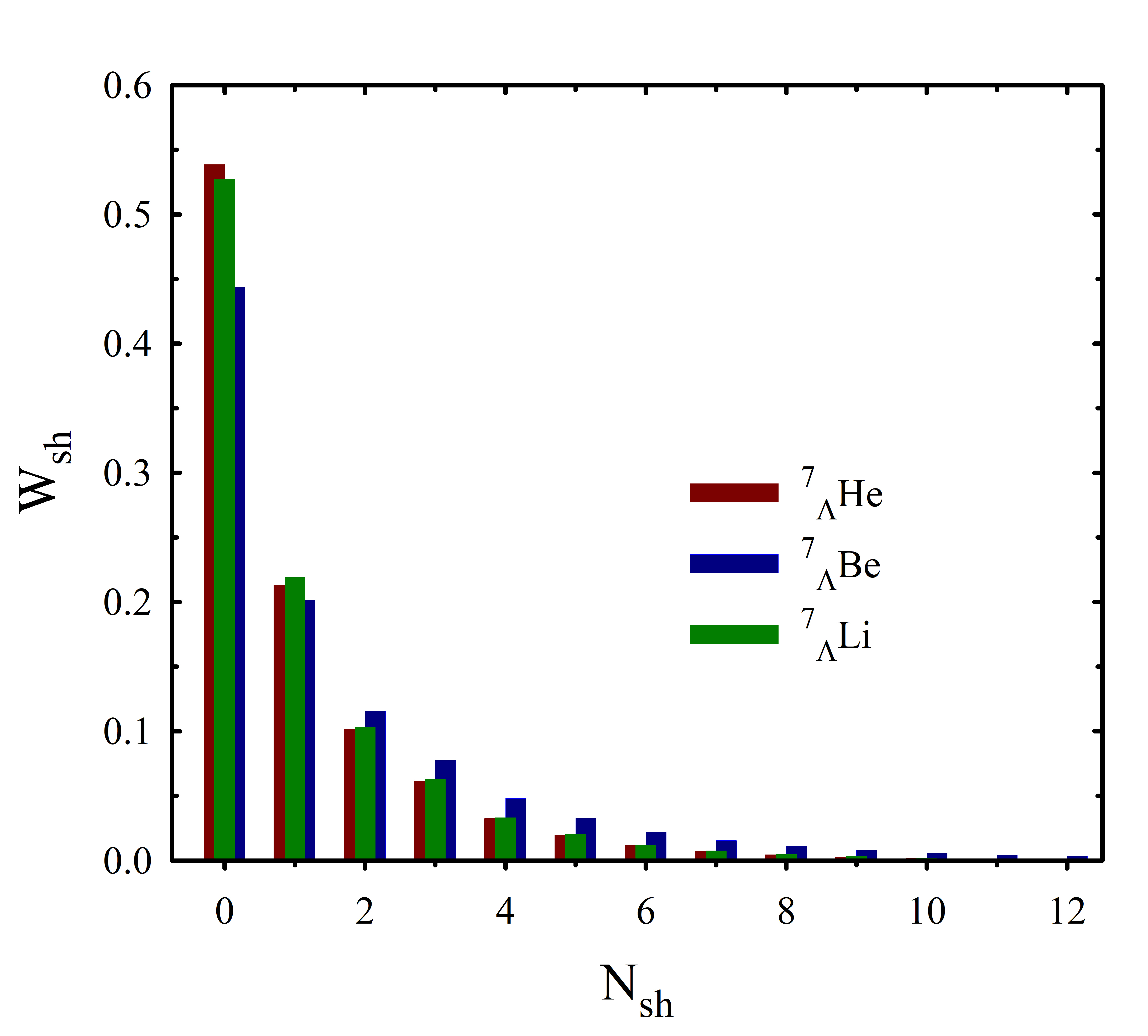}%
\caption{\label{Fig:Figure5}
Shell weights $W_{sh}$ in the wave functions of the ground states in
$_{\Lambda}^{7}$He, $_{\Lambda}^{7}$Be and $_{\Lambda}^{7}$Li. Results are obtained with the YNG-NF potential.}%
\end{center}
\end{figure}

In Fig. \ref{Fig:Figure5} we compare the structure of the ground
states of three hypernuclei   $_{\Lambda}^{7}$He, $_{\Lambda}^{7}$Be and
$_{\Lambda}^{7}$Li, which were obtained with the YNG-NF potential. One
notices that the oscillator functions belonging to the  lowest oscillator shell
$N_{sh}$=0 give the largest contribution to the ground state wave function. The
lower the bound state energy, the greater the contribution of the
shell $N_{sh}$=0. There is another correlation between the energy of the ground
state and the behavior of the weights $W_{sh}$. Indeed, the larger the bound state
energy (or in other words, the weaker is the bound state), the longer the tail
of the wave function in the oscillator representation. This is in full agreement with the
behavior of the bound state wave function in the coordinate space. We see that
the tail of the $_{\Lambda}^{7}$Be ground state is longer than the tails of the
more deeply bound states in $_{\Lambda}^{7}$Li and $_{\Lambda}^{7}$He.%

\begin{figure}[ht]
\begin{center}
\includegraphics[width=0.75\textwidth]{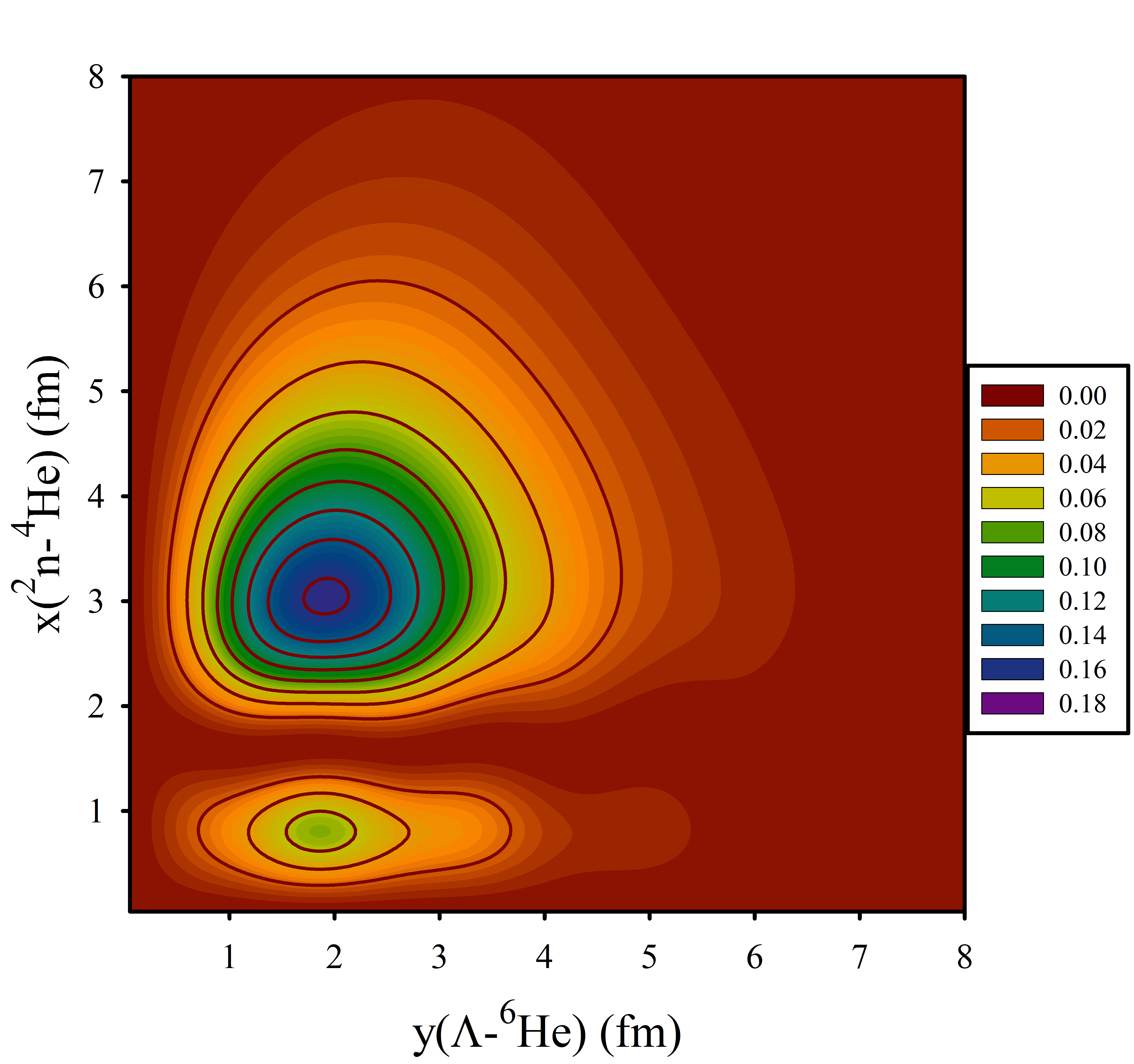}%
\caption{\label{Fig:Figure6}
Correlation function of the $_{\Lambda}^{7}$He ground state constructed with the YNG-NF potential.}%
\end{center}
\end{figure}

Fig. \ref{Fig:Figure6}, showing the correlation function for the ground state of
$_{\Lambda}^{7}$He, allows one to determine the most probable geometry of
three clusters in the coordinate space. The displayed correlation function has two
maxima.  The main maximum is found at x$\left(  ^{2}n-^{4}\text{He}\right)
=$3.02 fm and  y$\left(  \Lambda-^{6}\text{He}\right)  =$1.87 fm.  It means
that the lambda hyperon is located not far from the center of mass of $^{6}$He
and the latter is a fairly stretched nucleus. It
is interesting to note that the correlation functions of $_{\Lambda}^{7}$Li
and $_{\Lambda}^{7}$Be have the same form with slightly different coordinates
of the main maximum. For $_{\Lambda}^{7}$Be they are x$\left(  ^{2}%
p-^{4}\text{He}\right)  =$3.14 fm and  y$\left(  \Lambda-^{6}\text{Be}%
\right)  =$1.98 fm while for  $_{\Lambda}^{7}$Li \ they are equal to  x$\left(
d-^{4}\text{He}\right)  =$2.89 fm and  y$\left(  \Lambda-^{6}\text{Li}%
\right)  =$1.76 fm.%

Additional information on the structure of the bound states in the hypernuclei
$_{\Lambda}^{7}$He, $_{\Lambda}^{7}$Be and $_{\Lambda}^{7}$Li is presented in
Table \ref{Tab:7HHe7HBe7HLiAvDist}, where we show the
average distances between clusters. In the present paper, the quantity
$R_{B}$ represents the average distance of the two-cluster subsystem, consisting of
$^{4}$He and $d$ ($^2n$ or $^2p$), while $R_{\Lambda}$ determines the average
distance between the lambda hyperon and the two-cluster subsystem. We detected that
for relatively deeply bound states, i.e., the states with binding energy $E<$
-2.0 MeV,  the average sizes $R_{B}$ of binary subsystems are less than 4 fm
and the values of $ R_{\Lambda}$ are close to 3.0 fm. In Ref.
\cite{2025arXiv250901932K}, such the average distances $R_{\Lambda}$ have been
determined in a two-cluster model for lightest hypernuclei
comprised of the lambda hyperon and the s-shell nuclei, and it was, for example,
shown that the average distance between the lambda hyperon and alpha particle is
4.27 fm (the binding energy of $_{\Lambda}^{5}$He is -3.1 MeV) and between
the lambda hyperon and the deuteron is 9.73 fm (the binding energy of $_{\Lambda}^{3}%
$H is -0.16 MeV). The weakly bound states of $_{\Lambda}^{7}$He, $_{\Lambda
}^{7}$Be and $_{\Lambda}^{7} $Li have a large value of $R_{\Lambda}$, which exceeds the distance
4.27 fm between the lambda hyperon and the alpha particle, while the deeply bound states have
a fairly compact configuration relative to the lightest hypernuclei.%

\begin{table}[b]
\caption{\label{Tab:7HHe7HBe7HLiAvDist}
Energy of bound states, average distances between clusters
$R_{\Lambda}$, $R_B$  in $^{7}_{\Lambda}$He,   $^{7}_{\Lambda}$Be and  $^{7}_{\Lambda}$Li, determined
with the YNG-NF potential. Energy $E$ is given in MeV and average distances are in fm.}%
\begin{ruledtabular}
\begin{tabular}{cccccccccc}
\multicolumn{4}{c}{$_{\Lambda}^{7}$He} & \multicolumn{3}{c}{$_{\Lambda
}^{7}$Be} & \multicolumn{3}{c}{$_{\Lambda}^{7}$Li}\\
$J^{\pi}$ & $E$ & $R_{\Lambda}$ & $R_{B}$ & $E$ & $R_{\Lambda}$ & $R_{B}$ &
$E$ & $R_{\Lambda}$ & $R_{B}$\\\hline
1/2$^{+}$ & -6.05 & 2.96 & 3.68 & -4.31 & 3.05 & 3.96 & -7.10 & 2.90 &
3.33\\
3/2$^{+}$ & -3.34 & 2.75 & 3.42 & -1.57 & 2.99 & 4.36 & -6.96 & 2.88 &
3.34\\
5/2$^{+}$ & -2.73 & 2.95 & 3.77 & -1.05 & 3.29 & 5.09 & -4.85 & 2.65 &
2.88\\\
1/2$^{+}$ & -1.91 & 4.78 & 9.75 & -0.94 & 4.78 & 9.75 & -1.46 & 4.78 &
9.22\\
\end{tabular}
\end{ruledtabular}
\end{table}%
%

\subsection{AMHHB versus other methods}

To study the consistency of our model with other theoretical models, we compare the spectrum of bound states and their parameters, obtained in our model, with the results of other microscopic models. We pay much attention to those publications where energies of bound states are explicitly indicated along with their parameters, such as the mass, charge rms radii, and/or average distances between clusters.

First, we consider $_{\Lambda}^{7}$Li. In Table \ref{Tab:Spectr7HLiMethods} we
compare the spectrum of bound states in $_{\Lambda}^{7}$Li, obtained in our model
(AMHHB), with the results of the three-cluster  model
\cite{1983PThPh..70..189M} (Motoba1983) and
four-cluster models \cite{2006PhRvC..74e4312H} (Hiyama2006) and
\cite{2009PhRvC..80e4321H} (Hiyama2009). A three-cluster model Motoba1983, which is quite close to our model, was formulated in 
\cite{1983PThPh..70..189M} and applied to the study of several hypernuclei, including
$_{\Lambda}^{7}$Li. Both models consider the
same three-cluster structure for $_{\Lambda}^{7}$Li.  Contrary to our model,
the three-cluster model of Motoba et al \cite{1983PThPh..70..189M} uses the Orthogonality Condition Model
to describe a two-cluster subsystem $^{4}$He+d, with the one-Gaussian-type
potential for the central  and spin-orbit parts of the $^{4}$He+d interaction. In
this model  \cite{1983PThPh..70..189M}, the strength of the $N\Lambda$
interaction was selected to reproduce the experimental binding energy of the
lambda hyperon in  $_{\Lambda}^{5}$He. Both four-cluster models
\cite{2006PhRvC..74e4312H} (Hiyama2006) and \cite{2009PhRvC..80e4321H} (Hiyama2009)
also employ the orthogonality condition model to describe interactions of
valence nucleons and the lambda hyperon with the alpha particle. For the $N+\Lambda$
interaction, the authors use an effective  $N\Lambda$ potential, which simulates
the basic features of the Nijmegen potential NSC97f
\cite{1999PhRvC..59...21R}. Special remark concerning the Hiyama2009 model, in \cite{2009PhRvC..80e4321H}
the four-cluster is applied to study the spectrum of $_{\Lambda}^{7}$Li with
 the total isospin of $^{6}$Li  T=1, while in our model and in the Hiyama2006 model
the states with the total isospin T=0 for $^{6}$Li are considered. Note also
that the states of $^{6}$Li with the total isospin T=1 cannot be considered as the
$^{4}$He+d cluster structure and require a special case of the three-cluster
structure $^{4}$He+$p+n$.

In Ref. \cite{1994PThPS.117..361Y} (Yamamoto1994), another version of a three-cluster model, which is close to ours,
was realized to determine the spectrum of the bound states in $_{\Lambda}^{7}$Li.
The same set of  $\Lambda N$  potentials was used in Ref.
\cite{1994PThPS.117..361Y}. The Fermi momentum was selected to be $k_{F}$=
0.95 fm$^{-1}$ for all versions of the YNG potential. This value of $k_{F}$ is
very close to what we used for the NF version ($k_{F}$=0.9465) and the ND version
($k_{F}$=0.9460). With such a value of the parameter $k_{F}$, the
spectra of the bound states, determined with the NF and ND versions, are very
close to each other, however, with a different order of  the 1/2$^{+}$, 3/2$^{+}$,
5/2$^{+}$ and 7/2$^{+}$ states (see Fig. 4 in \cite{1994PThPS.117..361Y}).
Energies of these states, obtained with the NS version of the YNG potential,
are substantially different from those determined with the YNG-NF and YNG-ND
potentials.

\begin{figure}[ht]
\begin{center}
\includegraphics[width=0.75\textwidth]{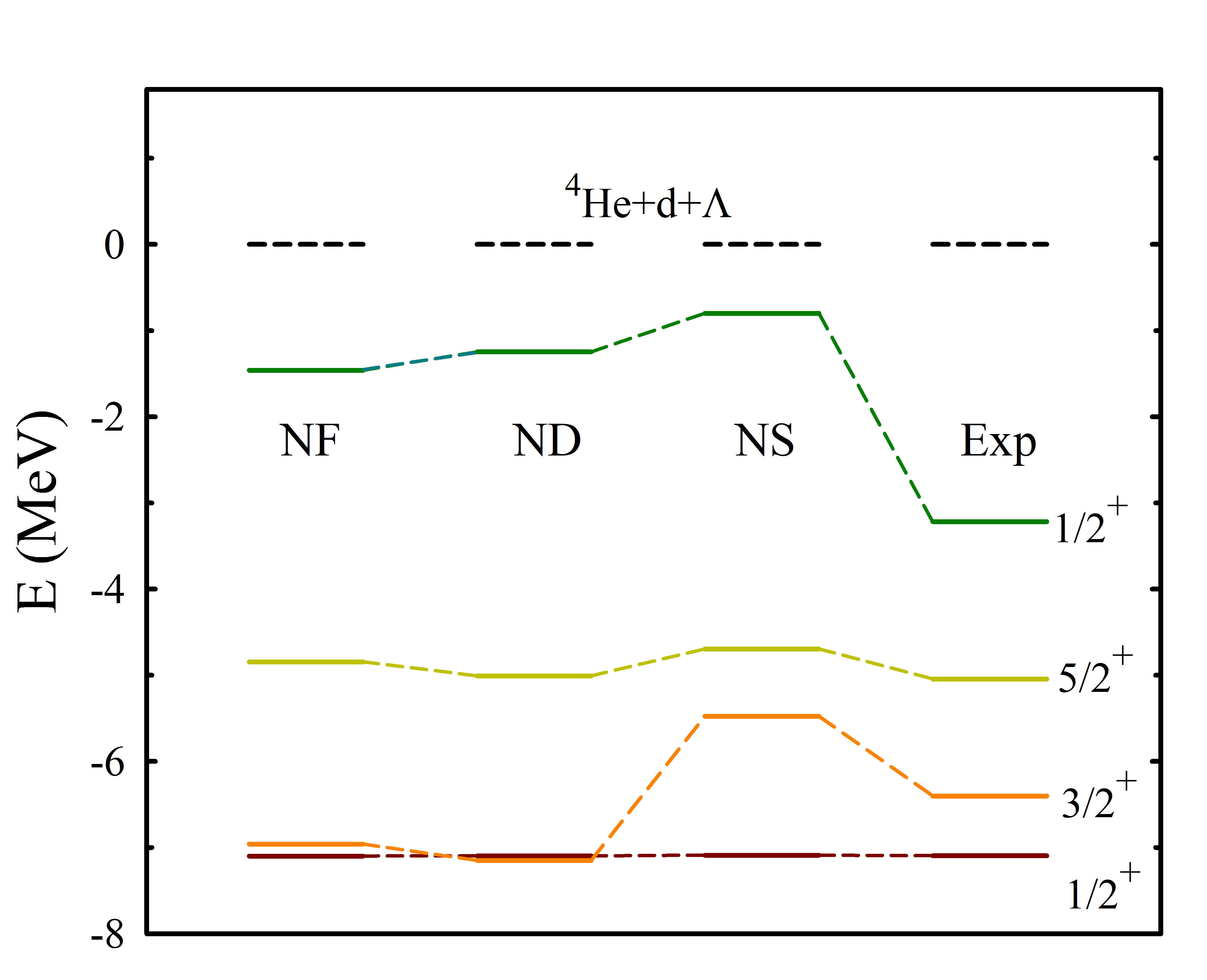}%
\caption{\label{Fig:Figure7}
Spectrum of bound states in $_{\Lambda}^{7}$Li, calculated with three
versions of the YNG potential. Experimental data are taken from the website
\cite{ChartHyperN2021}.}%
\end{center}
\end{figure}

Comparing our results (the spectrum of bound states) for $_{\Lambda}^{7}%
$Li, presented in Fig. \ref{Fig:Figure7}, with similar results,
obtained in  \cite{1994PThPS.117..361Y} and shown there in Fig.  4,  one notices a
certain resemblance between these results. In both models, the relative
position and order of the 1/2$^{+}$ and 3/2$^{+}$ strongly depend on the
shape of the YNG potential. In our model, the strongest impact of
different versions of YNG potential is observed for the 3/2$^{+}$ bound
states, and the minor impact is found for the 5/2$^{+}$ bound state. Fig.
\ref{Fig:Figure7} also demonstrates that our model with the selected
value of the parameters $k_{F}$ well reproduces the relative position of the
1/2$^{+}$ and 5/2$^{+}$ states.%

There is a certain agreement between the results of our three-cluster model and the three-cluster model, presented in \cite{1983PThPh..70..189M}. The mass rms radii of the bound states in  $_{\Lambda}^{7}$Li are quite close in both models, the average sizes of the two-cluster subsystem in our model are slightly less than in the Motoba1983 model. However, the average distances between the lambda hyperon and $^{6}$Li in our model are substantially larger than in the Motoba1983 model. %

\begin{table}[b]
\caption{\label{Tab:Spectr7HLiMethods}%
Spectrum of  $^{7}_{\Lambda}$Li  bound states determined with
different methods. The energy $E^{\ast}$ of the hypernucleus is measured from the corresponding four-channel threshold.}
\begin{ruledtabular}
\begin{tabular}{ccccccccc}
\multicolumn{5}{c}{AMHHB} & \multicolumn{4}{c}{Motoba1983,
\cite{1983PThPh..70..189M}}\\
$J^{\pi}$ & $E$ & $R_{m}$ & $R_{B}$ & $R_{\Lambda}$ & $E$ & $R_{m}$ & $R_{B}$
& $R_{\Lambda}$\\\hline
1/2$^{+}$ & -7.10 & 2.21 & 2.90 & 3.33 & -7.02 & 2.04 & 3.13 & 2.40\\
3/2$^{+}$ & -6.96 & 2.21 & 2.88 & 3.34 & -5.92 & 2.08 & 3.21 & 2.50\\
5/2$^{+}$ & -4.85 & 2.06 & 2.65 & 2.88 & -5.03 & 1.96 & 2.91 & 2.33\\\hline
\multicolumn{5}{c}{Hiyama2006, \cite{2006PhRvC..74e4312H}} &
\multicolumn{4}{c}{Hiyama2009, \cite{2009PhRvC..80e4321H}}\\
\multicolumn{1}{c}{$J^{\pi}$} & $E$ & $R_{m}$ & \multicolumn{1}{c}{$R_{B}%
$} &  & $E^{\ast}$ & $R_{m}$ & $R_{B}$ & $R_{\Lambda}$\\\hline
\multicolumn{1}{c}{1/2$^{+}$} & \multicolumn{1}{c}{-5.59} &  &  &  &
-5.40 & 3.74 &  & \\
\multicolumn{1}{c}{3/2$^{+}$} & \multicolumn{1}{c}{-4.90} &  &  &  &
-3.75 & 3.92 &  & \\
\multicolumn{1}{c}{5/2$^{+}$} & \multicolumn{1}{c}{-3.33} &  &  &  &
-3.66 & 3.96 &  & \\
\end{tabular}
\end{ruledtabular}
\end{table}%

The spectra of the bound states in $_{\Lambda}^{7}$He and $_{\Lambda}^{7}$Be, investigated by two
different theoretical methods, are shown in Table \ref{Tab:7HHe7HBeMethods}. As
was pointed out above, in Refs. \cite{2009PhRvC..80e4321H}, 
\cite{2015PhRvC..91e4316H}, \cite{2023PhRvC.107e4302M} they have been determined within a four-cluster model
and thus energies of the bound states are measured from the four-cluster
threshold, which we marked them as $E^{\ast}$ in Table
\ref{Tab:7HHe7HBeMethods}. In Ref. \cite{2009PhRvC..80e4321H}, the average
distances between clusters have been determined, but they are of different
types with respect to ours. They are $R_{\alpha-N}$\ and $R_{\alpha-\Lambda}%
$,\ and represent the average distance of the valence nucleon with respect to
the alpha particle and the average distance of the lambda hyperon with respect
to the alpha particle, respectively. Recall that in the present work, we
calculated the average size $R_{B}$ of two-cluster subsystems $^{6}$He, $^{6}%
$Li, and $^{6}$Be, considered as the two-cluster systems. 

The Gamow shell model
 has been applied in Ref. \cite{2025PhLB..86839708L} (Li2025) to study a list of
neutron-rich He hyper-isotopes, among them is $_{\Lambda}^{7}$He. The
neutron rms radius of the $_{\Lambda}^{7}$He ground state,
determined in \cite{2025PhLB..86839708L}, equals  $R_{n}$ = 2.531 fm, which is
slightly larger than the rms radius $R_{m}$=2.35 fm, determined
in our model. The large value of $R_{n}$ indicates that there is a neutron
halo in $_{\Lambda}^{7}$He.

As we see, in $_{\Lambda}^{7}$He, the 3/2$^{+}$ excited state lies above the ground state at
energy $E$=2.71 MeV in our model, which is greater than in  the Li2025  model \cite{2025PhLB..86839708L} ($E$=1.51
MeV) and in a four-cluster model Hiyama2009 ($E$=1.66 MeV). For the 5/2$^{+}$
excited state, our model places it at $E$=3.32 MeV, the Li2025 model places it
not very far from our model at  $E$=2.68 MeV and the four-cluster model
Hiayama2009 \cite{2009PhRvC..80e4321H} creates this state at $E$=1.75 MeV. In our opinion, such a difference of the relative position of bound states in $_{\Lambda}^{7}$He
originates mainly from the shapes of the nucleon-nucleon and hyperon-nucleon
potentials. Obviously, peculiarities of the models applied also affect the
spectrum of the bound states.

\begin{table}[b]
\caption{\label{Tab:7HHe7HBeMethods}
Energy of bound states, average distances between clusters
$R_{\Lambda}$, $R_B$ obtained in $^{7}_{\Lambda}$He  and  $^{7}_{\Lambda}$Be
with the YNG-NF potential. The energy $E^{\ast}$ of the hypernuclei is measured from the four-channel threshold.}%
\begin{ruledtabular}
\begin{tabular}{cccccccccccc}
$_{\Lambda}^{A}$Z & \multicolumn{5}{c}{AMHHB} &
\multicolumn{3}{c}{Hiyama2009, \cite{2009PhRvC..80e4321H}} &
\multicolumn{3}{c}{Li2025 , \cite{2025PhLB..86839708L}}\\
& $J^{\pi}$ & $E$ & $R_{\Lambda}$ & $R_{B}$ & $R_{m}$ & $E^{\ast}$ &
$R_{\alpha-\Lambda}$ & $R_{\alpha-N}$ & $E^{\ast}$ & $R_{n}$ & $R_{\Lambda}%
$\\\hline
$_{\Lambda}^{7}$He & 1/2$^{+}$ & -6.05 & 2.96 & 3.68 & 2.35 & -6.39 & 2.81 &
3.66 & -6.06 & 2.531 & 2.447\\
& 3/2$^{+}$ & -3.34 & 2.75 & 3.42 & 2.35 & -4.73 & 2.79 & 3.80 & -4.55 &  &
\\
& 5/2$^{+}$ & -2.73 & 2.95 & 3.77 & 2.38 & -4.65 & 2.78 & 3.83 & -3.38 &  &
\\\hline
$_{\Lambda}^{7}$Be & $J^{\pi}$ & $E$ & $R_{\Lambda}$ & $R_{B}$ & $R_{m}$ &
$E^{\ast}$ & $R_{\alpha-\Lambda}$ & $R_{\alpha-N}$ &  &  & \\\hline
& 1/2$^{+}$ & -4.31 & 3.05 & 3.96 & 2.44 & -4.42 & 2.83 & 3.84 &  &  &
\\
& 3/2$^{+}$ & -1.57 & 2.99 & 4.36 & 2.56 &  &  &  &  &  & \\
& 5/2$^{+}$ & -1.05 & 3.29 & 5.09 & 2.87 &  &  &  &  &  & \\
\end{tabular}
\end{ruledtabular}
\end{table}%

\subsection{Three-cluster resonances}

Now we turn our attention to  resonance states embedded
in the three-cluster continuum of hypernuclei $_{\Lambda}^{7}$He, $_{\Lambda
}^{7}$Be and $_{\Lambda}^{7}$Li, and to dominant channels of their decay.

\begin{figure}[ht]
\begin{center}
\includegraphics[width=0.75\textwidth]{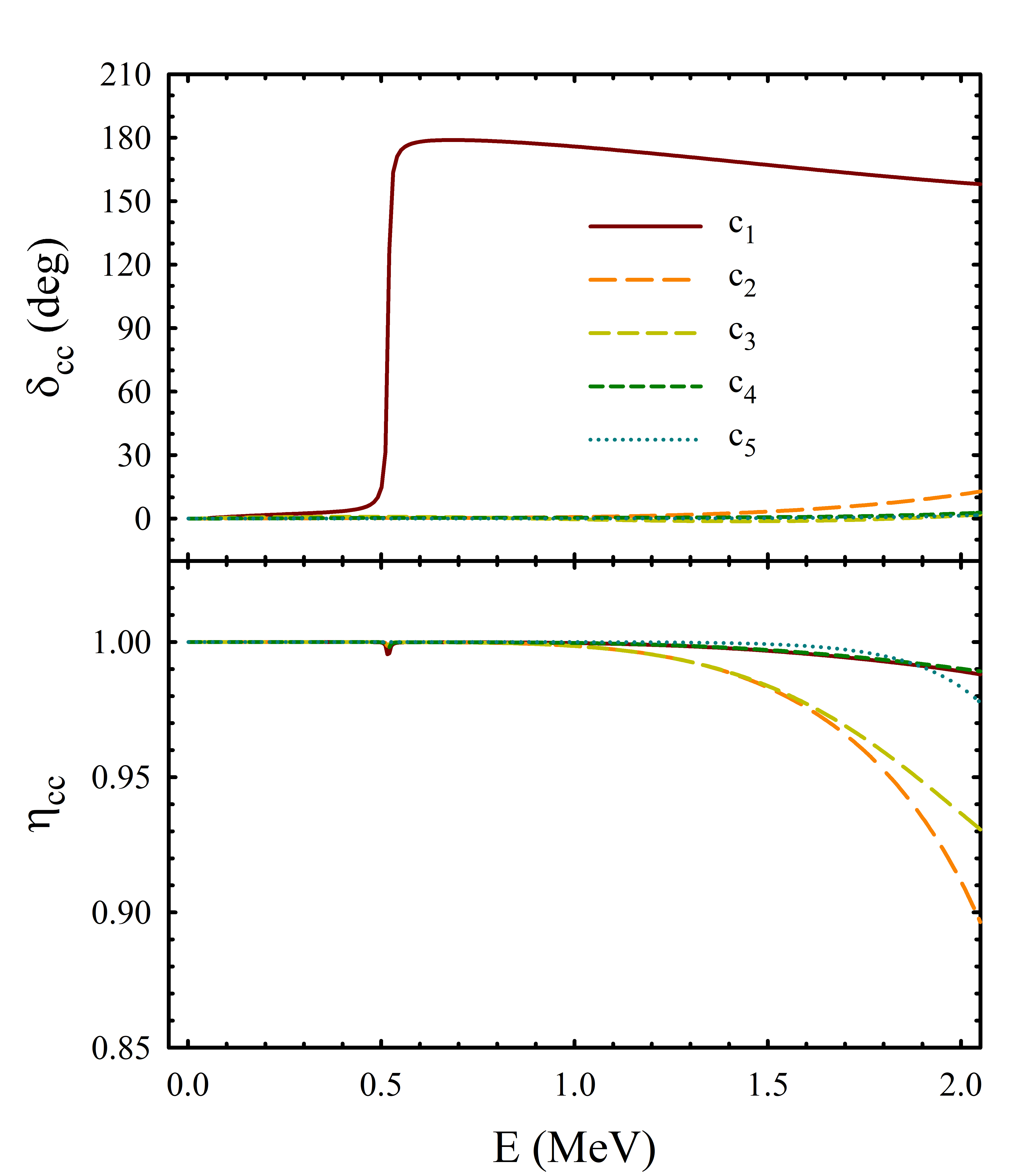}%
\caption{\label{Fig:Figure8}
Diagonal phase shifts $\delta_{cc}$ and inelastic parameters
$\eta_{cc}$ as a function of energy, determined with the YNG-NF potential for
the 5/2$^{+}$ state in $_{\Lambda}^{7}$Li.}%
\end{center}
\end{figure}

Before displaying parameters of the resonance states found, we demonstrate how
these resonance states manifest themselves in the scattering S-matrix, more
precisely, in the diagonal phase shifts $\delta_{cc}$ and diagonal inelastic
parameters $\eta_{cc}$. They are determined in Eq. (\ref{eq:S005}) and related
to the channel $c$ with the quantum numbers $c=\left\{  K,l_{y},l_{x}%
,L,S\right\}  $ and determine cross sections of the elastic 3-to-3 scattering.
 In Fig. \ref{Fig:Figure8} phase shifts and inelastic parameters are
shown for the 5/2$^{+}$ state in $_{\Lambda}^{7}$Li. These quantities are
demonstrated for five channels: $c_{1}=\{2,0,2,2,1/2\}$, $ c_{2}%
=\{2,1,1,2,1/2\}$, $ c_{3}=\{2,2,0,2,1/2\}$, $c_{4}=\{4,0,2,2,1/2\}$,
$c_{5}=\{4,1,1,2,1/2\}$. The phase shift of the first channel exhibits
resonance behavior at energy $E$ around 0.5 MeV, where the phase shift
increases rapidly on 180$^{\circ}$. All inelastic parameters displayed 
$\eta_{cc}$ have a tiny minimum at the energy $E$ =0.519 MeV, and the deepest
minimum is for the first channel. By using an algorithm, presented by Eq.
(\ref{eq:S008}), we found the energy of the 5/2$^{+}$ resonance state
$E$ =0.519 MeV and its width $\Gamma$ =8.35 keV.

\begin{figure}[ht]
\begin{center}
\includegraphics[width=0.75\textwidth]{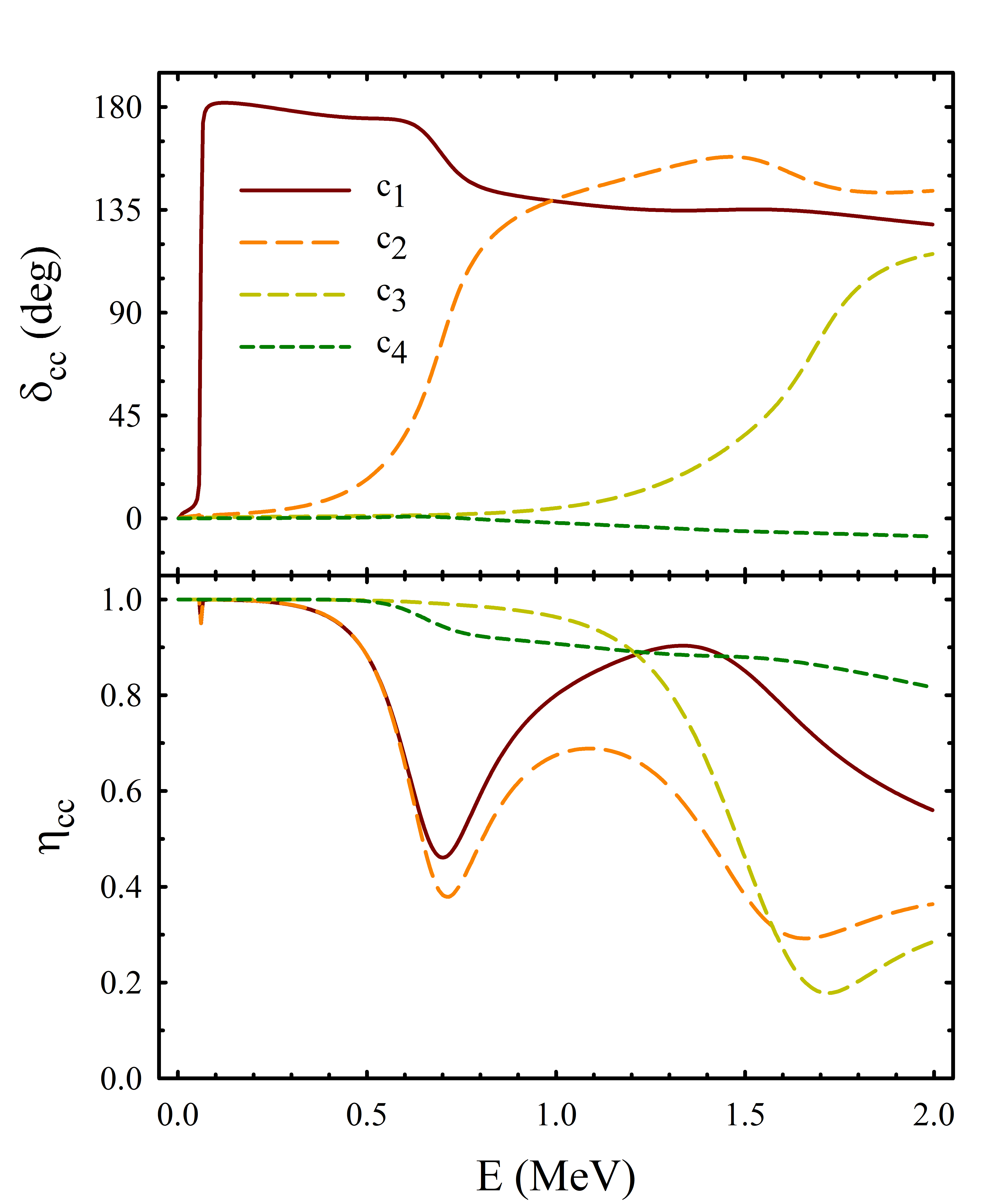}%
\caption{\label{Fig:Figure9}
Phase shifts and inelastic parameters for 1/2$^{-}$ state in
$_{\Lambda}^{7}$He, calculated with the YNG-NF potential.}%
\end{center}
\end{figure}

Fig. \ref{Fig:Figure8} shows a typical behavior of the scattering
parameters for the very narrow resonance states, which are observed in
hypernuclei $_{\Lambda}^{7}$He, $_{\Lambda}^{7}$Be and $_{\Lambda}^{7}$Li. To
confirm this, in Fig. \ref{Fig:Figure9} we show phase shifts and
inelastic parameters for the 1/2$^{-}$ state in $_{\Lambda}^{7}$He. This
figure reveals three resonance states. The first  is a very narrow
resonance state and lies close to the three-cluster threshold (E$\approx$ 0.06
MeV). This resonance state manifests itself in the first channel $c_{1}%
$=$\left\{  1,1,0,1,1/2\right\}  $. The second and  third resonance states
manifest themselves in the second ($c_{2}$=$\left\{  1,0,1,1,1/2\right\}  $)
and third ($c_{3}$=$\left\{  3,1,0,1,1/2\right\}  $) channels, respectively.
Phase shifts demonstrate that the second resonance state is narrower than the
third state, since the second phase shift is growing faster than the third
one. One can see that all resonance states make a local minimum of the
inelastic parameters as a function of energy. Using the Breit-Wigner
approximation for eigenphase shifts or relations (\ref{eq:S008}), we obtain
parameters of three resonance states: $E$=0.061 MeV, $\Gamma$=2.15 keV; $E
$=0.651 MeV, $\Gamma$=362.76 keV; $E$=1.453 MeV, $\Gamma$=912.77 keV.%

In Tables \ref{Tab:RS7HHE3potON}, \ref{Tab:Resons7HLi3PotKF},
\ref{Tab:RS7HBE3potONN} we collected parameters of three-cluster resonance
states in hypernuclei $_{\Lambda}^{7}$He, $_{\Lambda}^{7}$Li and $_{\Lambda
}^{7}$Be. We present results obtained with three versions of the YNG
potential. Turn our attention to the resonance states in $_{\Lambda}^{7}$He.
One can see that the number of resonance states discovered and their
parameters depend on the shape of the N$\Lambda$ potential. The YNG-NF and
YNG-ND potentials create the same number of resonance states, and some of
these resonances have close parameters. In contrast to those potentials, the
YNG-NS potential does not generate 3/2$^{+}$ and 5/2$^{-}$ resonance states.
It seems that the combination of the attractive part of the intercluster interaction and the
centrifugal barrier in these states does not create a favorable situation for
accommodation of a resonance state. One can also see that our model reveals
several very narrow resonance states (three resonance states with the YNG-NF
and YNG-ND potentials, and two resonance states with the YNG-NS potential) with the
 total width less than 15 keV, they are of negative parity, two of them lie
close to the three-cluster threshold $^{4}$He+$^2n$+$\Lambda$.

By using the same input parameters, we calculated the spectrum of resonance states
in $_{\Lambda}^{7}$Be which are shown in Table \ref{Tab:RS7HBE3potONN}. The
Coulomb interaction reduces the number of observed resonance states in the
three-cluster continuum of $_{\Lambda}^{7}$Be compared to their number in
$_{\Lambda}^{7}$He. Later, we will discuss in detail the effects of the Coulomb
interaction in mirror hypernuclei $_{\Lambda}^{7}$He and $_{\Lambda}^{7}$Be.
One can see that all three potentials form two fairly narrow resonance
states, their total widths are less than 41 keV and energies are varied
between 1.06 and 1.19 MeV.%

\begin{table}[b]
\caption{\label{Tab:RS7HHE3potON}
Spectrum of resonance states in $^7_{\Lambda}$He, obtained with
three $N\Lambda$ potentials. Energy is MeV and width is in keV.}%
\begin{ruledtabular}
\begin{tabular}{ccccccccc}
\multicolumn{3}{c}{YNG-NF} & \multicolumn{3}{c}{YNG-ND} & 
\multicolumn{3}{c}{YNG-NS} \\ 
\multicolumn{3}{c}{NF, $k_{F}$=0.9470} & \multicolumn{3}{c}{ND, $k_{F}$%
=0.9530} & \multicolumn{3}{c}{NS, $k_{F}$= 0.9410} \\ 
$J^{\pi }$ & $E$ & $\Gamma $ & $J^{\pi }$ & $E$ & $\Gamma $
& $J^{\pi }$ & $E$ & $\Gamma $ \\ \hline
3/2$^{-}$ & 0.043 & 1.09 & 3/2$^{-}$ & 0.027 & 14.49 & 3/2$^{-}$ & 0.057 & 
2.14 \\ 
1/2$^{-}$ & 0.061 & 2.15 & 1/2$^{-}$ & 0.048 & 1.27 & 1/2$^{-}$ & 0.057 & 
2.43 \\ 
5/2$^{-}$ & 0.493 & 5.09 & 5/2$^{-}$ & 0.490 & 5.36 & 5/2$^{-}$ & 0.606 & 
11.77 \\ 
3/2$^{-}$ & 0.622 & 350.57 & 5/2$^{+}$ & 0.655 & 194.17 & 5/2$^{+}$ & 0.657
& 195.95 \\ 
1/2$^{-}$ & 0.651 & 362.76 & 3/2$^{+}$ & 0.656 & 195.33 & 3/2$^{+}$ & 0.657
& 195.95 \\ 
3/2$^{+}$ & 0.656 & 196.46 & 3/2$^{-}$ & 0.658 & 384.19 & 3/2$^{-}$ & 0.692
& 433.60 \\ 
5/2$^{+}$ & 0.656 & 195.17 & 1/2$^{-}$ & 0.692 & 392.03 & 1/2$^{-}$ & 0.692
& 433.60 \\ 
1/2$^{+}$ & 1.226 & 689.83 & 1/2$^{+}$ & 1.223 & 651.01 & 1/2$^{+}$ & 1.280
& 731.99 \\ 
3/2$^{-}$ & 1.401 & 754.43 & 3/2$^{-}$ & 1.347 & 655.71 & 1/2$^{+}$ & 1.280
& 731.99 \\ 
5/2$^{-}$ & 1.417 & 620.56 & 1/2$^{-}$ & 1.407 & 818.80 & 5/2$^{-}$ & 1.419
& 618.48 \\ 
1/2$^{-}$ & 1.453 & 912.77 & 5/2$^{-}$ & 1.417 & 619.35 & 1/2$^{-}$ & 1.456
& 842.36 \\ 
1/2$^{+}$ & 2.087 & 4060.68 & 1/2$^{+}$ & 2.068 & 393.85 & 3/2$^{-}$ & 1.456
& 842.36 \\ 
5/2$^{+}$ & 2.194 & 797.54 & 5/2$^{+}$ & 2.233 & 817.61 & 5/2$^{-}$ & 1.419
& 618.48 \\ 
3/2$^{+}$ & 2.271 & 846.41 & 3/2$^{+}$ & 2.294 & 861.69 &  3/2$^{+}$ & 2.296 & 877.52\\ 
 &  &  &  &  &  & 3/2$^{+}$ & 2.296 & 877.52 \\ 
 &  &  &  &  &  & 5/2$^{+}$ & 2.296 & 877.52 \\ 
\end{tabular}
\end{ruledtabular}
\end{table}%

\begin{figure}[ht]
\begin{center}
\includegraphics[width=0.75\textwidth]{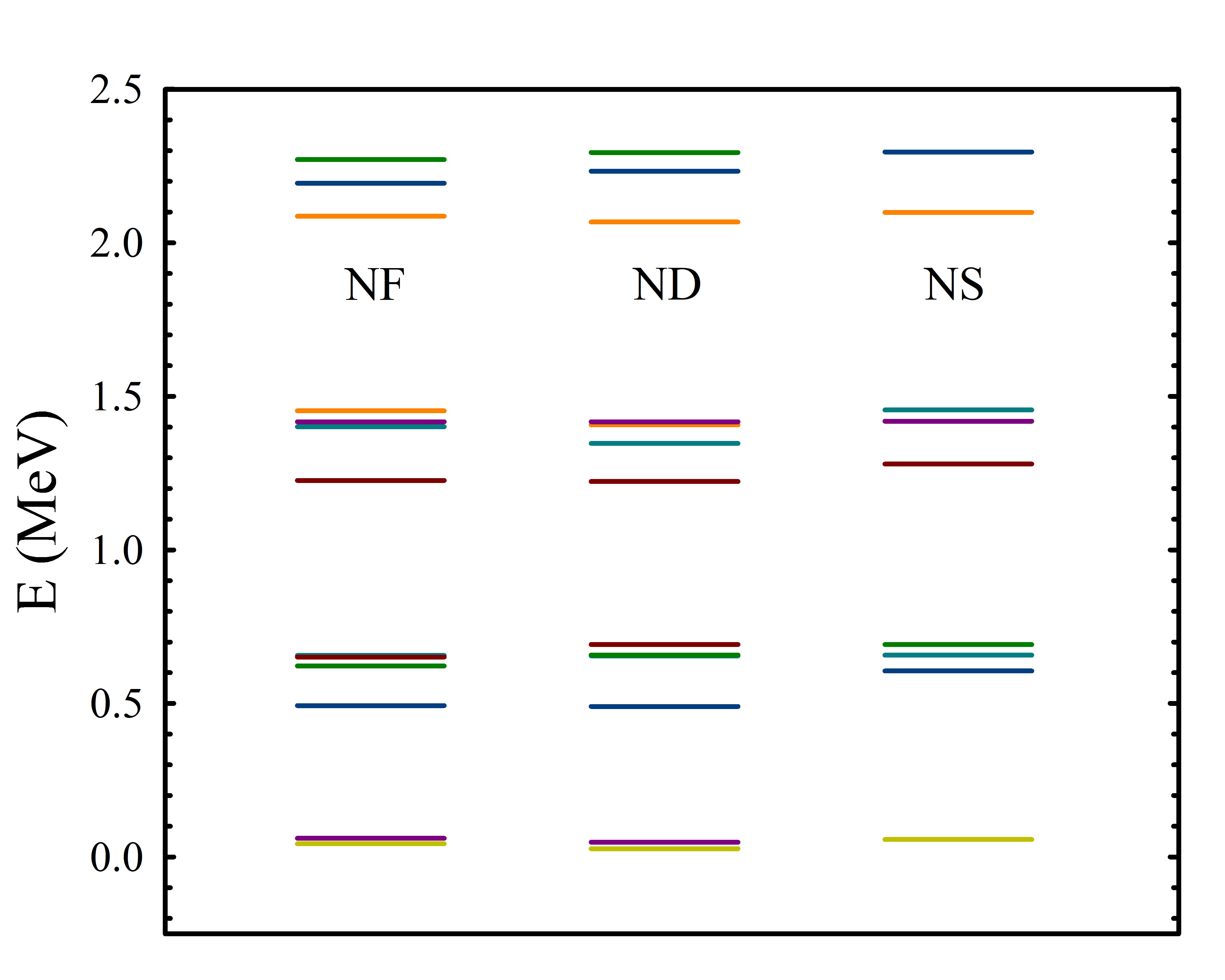}%
\caption{\label{Fig:Figure10}%
Spectra of the three-cluster resonance states in $_{\Lambda}^{7}$He
determined with the NF, ND and NS versions of the YNG potential.}%
\end{center}
\end{figure}

\begin{figure}[ht]
\begin{center}
\includegraphics[width=0.75\textwidth]{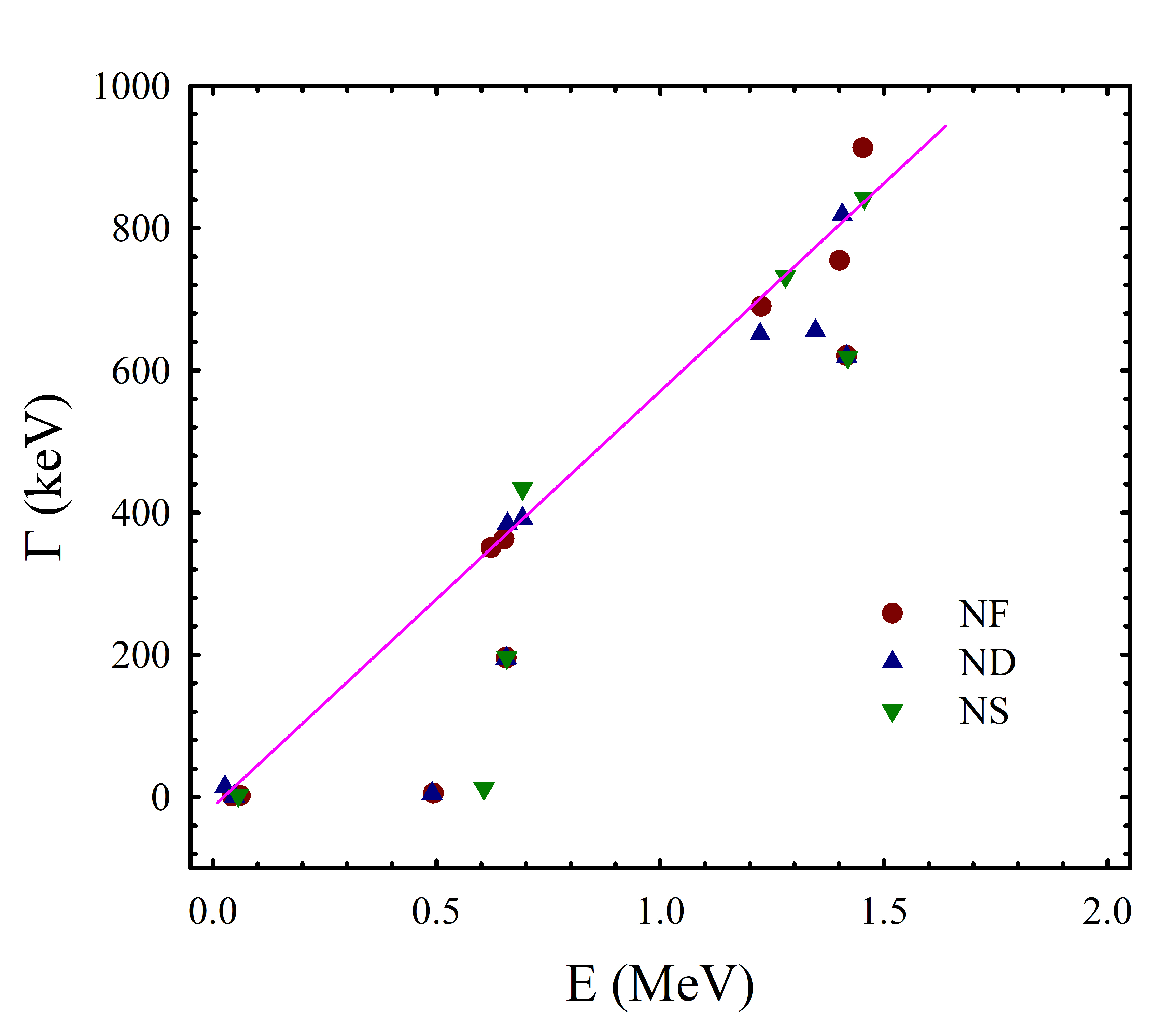}%
\caption{\label{Fig:Figure11}
Correlation between energies and widths of three-cluster resonance
states in $_{\Lambda}^{7}$He, determined with different versions of the YNG
potential.}%
\end{center}
\end{figure}

In Figs. \ref{Fig:Figure10} and \ref{Fig:Figure11}, we visualize the
results shown in Table \ref{Tab:RS7HHE3potON}. In Fig.
\ref{Fig:Figure10} we compare the position of  the resonance states in
$_{\Lambda}^{7}$He obtained with three versions of the YNG potential. Fig.
\ref{Fig:Figure10} reveals a shell-like structure of three-cluster
resonance states in $_{\Lambda}^{7}$He, where the resonance states are separated
into several groups with close energies and fairly large gaps between these
groups. Fig. \ref{Fig:Figure11} shows both the energies and total widths of the
resonance states in $_{\Lambda}^{7}$He. One can see that there is a certain correlation between the energy and width of the resonance states, which are determined
with three versions of the YNG potential. Indeed, the majority of the
displayed resonances (12 of the total 48 resonances) form a straight line. We assume
that this correlation is due to the peculiarities of the used versions of the
YNG potential and the specific features of the three-cluster model of the
hypernucleus $_{\Lambda}^{7}$He.%

\begin{figure}[ht]
\begin{center}
\includegraphics[width=0.75\textwidth]{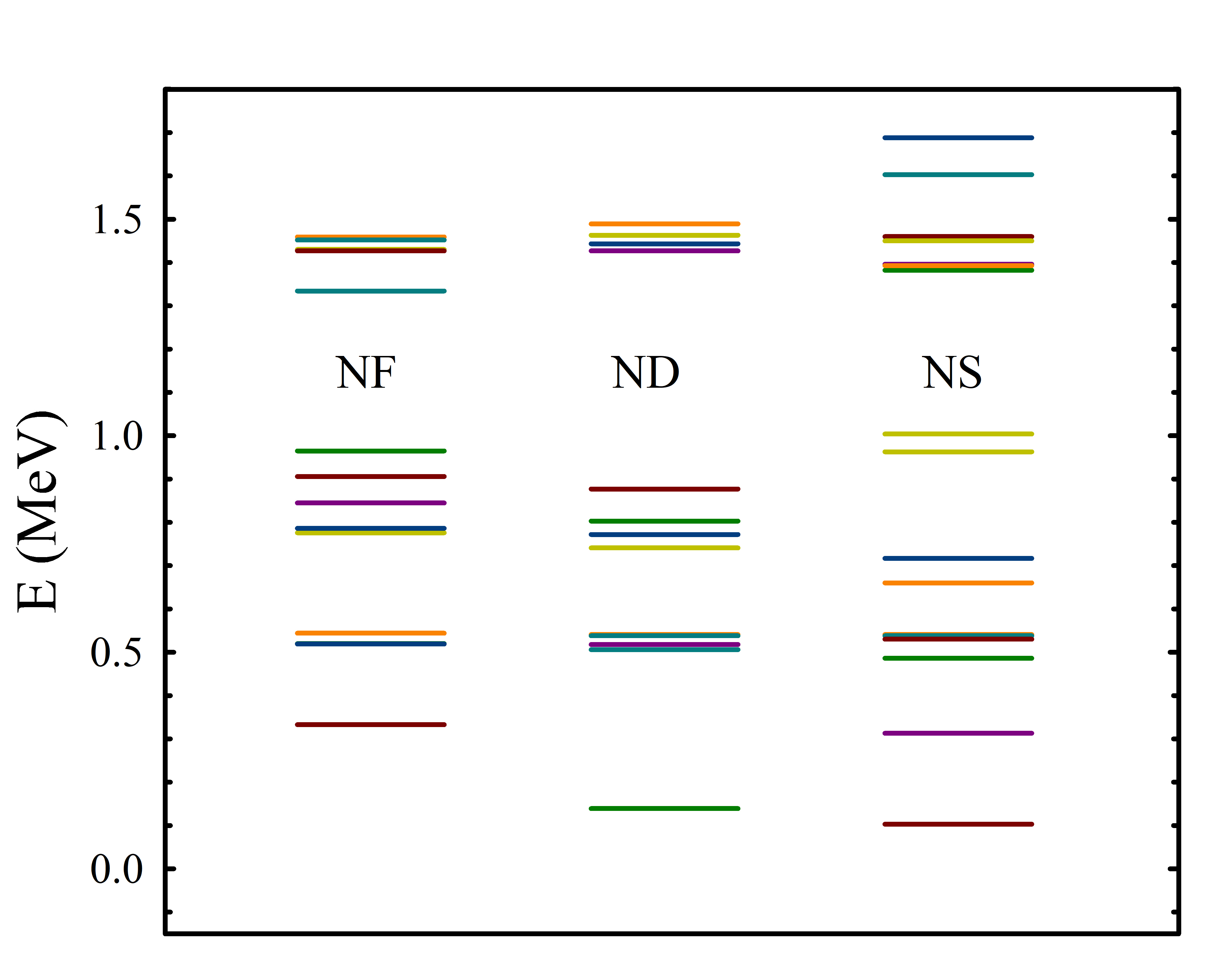}%
\caption{\label{Fig:Figure12}%
Spectra of resonance states in the three-cluster continuum of $_{\Lambda
}^{7}$Li, determined with three versions of the YNG potential.}%
\end{center}
\end{figure}

And now we turn our attention to the resonance states in the hypernucleus
$_{\Lambda}^{7}$Li. Their parameters are shown in Table
\ref{Tab:Resons7HLi3PotKF} and the relative position of resonance states is demonstrated in Fig. \ref{Fig:Figure12}.
 Similar to the case of  $_{\Lambda}^{7}$He, three versions of the YNG potential generate a
shell-like structure of resonance spectra, however, it is not as bright 
as for the YNG-NS potential. Comparing spectra of the resonance
states in $_{\Lambda}^{7}$He ( Fig. \ref{Fig:Figure10}) and $_{\Lambda}^{7}$Li (Fig.
\ref{Fig:Figure12}), we see that the shape of the YNG potentials has a
stronger influence on the position of the resonance states in $_{\Lambda}%
^{7}$Li than in $_{\Lambda}^{7}$He.

Analyzing the parameters of the resonance states in  $_{\Lambda}%
^{7}$Li, we see that the YNG-NF and YNG-ND potentials form four very narrow resonance states with total width $\Gamma<$
15 keV, and five very narrow resonance states with  width $\Gamma<$
11 keV are created by the YNG-NS potential. The energies of the
narrowest resonance states and their total angular momentum and parity are
affected by the shape of the YNG potentials. However, there are some
narrowest resonance states with the same angular momentum $J$ and parity $\pi
$, and their parameters are very close. They are, for example, the 5/2$^{+}$
resonance states with  energy $E$ = 0.519 MeV and total width $\Gamma$ =8 keV.
There are also other, wider resonance states with very close parameters.%

The narrowest resonance state in $_{\Lambda}^{7}$Li, obtained with the YNG-NF
potential, has parameters $E$ =0.333 MeV and $\Gamma$ = 1.5 keV and is
formed in the 1/2$^{+}$ state. The YNG-NS potential creates the narrowest potential in  the 5/2$^{+}$ state  with very close parameters: $E$ =0.345 MeV and $\Gamma$ = 2.2 keV. The YNG-ND potential creates the narrowest
resonance state with $J^{\pi}$=5/2$^{+}$, with the energy $E$=0.518 MeV and
width $\Gamma$= 8.3 keV. Almost the same parameters possess the 5/2$^{+}$
resonance state formed by the YNG-NF potential.

\begin{table}[b]
\caption{\label{Tab:Resons7HLi3PotKF}
Parameters of resonance states found in the three-cluster continuum of
$^7_{\Lambda}$Li. Energy is in MeV and width is in keV.}%
\begin{ruledtabular}
\begin{tabular}{ccccccccc}
\multicolumn{3}{c}{YNG-NF} & \multicolumn{3}{c}{YNG-ND} & 
\multicolumn{3}{c}{YNG-NS} \\ 
\multicolumn{3}{c}{$k_{F}$=0.9465} & \multicolumn{3}{c}{$k_{F}$=0.9460} & 
\multicolumn{3}{c}{$k_{F}$=1.0440} \\ 
$J^{\pi }$ & $E$ & $\Gamma $ & $J^{\pi }$ & $E$ & $\Gamma $ & $J^{\pi }$ & $E$ & $\Gamma $ \\ \hline
1/2$^{+}$ & 0.333 & 1.54 & 5/2$^{-}$ & 0.506 & 53.19 & 1/2$^{+}$ & 0.139 & 
388.93 \\ 
5/2$^{+}$ & 0.519 & 8.35 & 5/2$^{+}$ & 0.518 & 8.26 & 5/2$^{+}$ & 0.345 & 
2.16 \\ 
3/2$^{+}$ & 0.520 & 8.43 & 3/2$^{+}$ & 0.538 & 10.34 & 3/2$^{+}$ & 0.486 & 
10.15 \\ 
1/2$^{+}$ & 0.544 & 11.12 & 1/2$^{+}$ & 0.541 & 10.66 & 5/2$^{+}$ & 0.519 & 
8.36 \\ 
3/2$^{-}$ & 0.776 & 252.18 & 3/2$^{-}$ & 0.741 & 207.15 & 3/2$^{+}$ & 0.538
& 10.31 \\ 
5/2$^{-}$ & 0.786 & 264.26 & 3/2$^{-}$ & 0.803 & 14.71 & 3/2$^{-}$ & 0.717 & 
130.37 \\ 
5/2$^{-}$ & 0.845 & 17.25 & 1/2$^{-}$ & 0.877 & 350.81 & 1/2$^{-}$ & 0.963 & 
258.76 \\ 
1/2$^{-}$ & 0.906 & 386.64 & 3/2$^{-}$ & 1.427 & 164.31 & 1/2$^{+}$ & 1.198
& 246.68 \\ 
3/2$^{-}$ & 0.965 & 432.09 & 3/2$^{+}$ & 1.443 & 526.39 & 1/2$^{-}$ & 1.349
& 825.22 \\ 
1/2$^{+}$ & 1.430 & 463.13 & 1/2$^{+}$ & 1.463 & 553.83 & 5/2$^{-}$ & 1.393
& 151.64 \\ 
3/2$^{-}$ & 1.452 & 184.85 & 1/2$^{-}$ & 1.489 & 212.93 & 3/2$^{-}$ & 1.396
& 151.51 \\ 
1/2$^{-}$ & 1.459 & 192.70 &  &  &  & 1/2$^{-}$ & 1.603 & 555.46 \\ 
\end{tabular}
\end{ruledtabular}
\end{table}%

It is interesting to note that the resonance structure of $_{\Lambda}^{7}$He has
been investigated in Ref. \cite{2015PhRvC..91e4316H} within a four-cluster
model. Two resonance states are determined with the following parameters:
 $E\left(  5/2^{+}\right)  $ =0.07 MeV, $\Gamma\left(  5/2^{+}\right)  $
=1.01 MeV, and $E\left(  3/2^{+}\right)  $ =0.03 MeV, $\Gamma\left(
3/2^{+}\right)  $ =1.13 MeV. Energy is measured from the four-cluster
threshold $^{4}$He+n+n+$\Lambda$.%

\begin{table}[b]
\caption{\label{Tab:RS7HBE3potONN}
Spectrum of resonance states in $^7_{\Lambda}$Be, obtained with
three $N\Lambda$ potentials. Energy is MeV and width is in keV.}%
\begin{ruledtabular}
\begin{tabular}{ccccccccc}
\multicolumn{3}{c}{NF} & \multicolumn{3}{c}{ND} & \multicolumn{3}{c}{NS
} \\ 
\multicolumn{3}{c}{NF, $k_{F}$=0.9470} & \multicolumn{3}{c}{ND, $k_{F}$%
=0.9530} & \multicolumn{3}{c}{NS, $k_{F}$= 0.9410} \\ 
$J^{\pi }$ & $E$ & $\Gamma $ &  & $E$ & $\Gamma $ & $%
J^{\pi }$ & $E$ & $\Gamma $ \\ \hline
5/2$^{-}$ & 1.067 & 25.19 & 5/2$^{-}$ & 1.055 & 25.29 & 5/2$^{-}$ & 1.182 & 
40.73 \\ 
1/2$^{+}$ & 1.079 & 969.05 & 1/2$^{+}$ & 1.094 & 889.71 & 1/2$^{+}$ & 1.075
& 960.15 \\ 
3/2$^{-}$ & 1.246 & 477.27 & 3/2$^{-}$ & 1.235 & 457.07 & 3/2$^{-}$ & 1.257
& 513.82 \\ 
1/2$^{-}$ & 1.257 & 518.33 & 1/2$^{-}$ & 1.247 & 498.75 & 1/2$^{-}$ & 1.257
& 513.82 \\ 
3/2$^{-}$ & 1.474 & 508.85 & 3/2$^{-}$ & 1.535 & 490.81 & 3/2$^{-}$ & 1.553
& 565.55 \\ 
1/2$^{-}$ & 1.490 & 500.94 & 1/2$^{-}$ & 1.550 & 4812.87 & 1/2$^{-}$ & 1.553
& 565.55 \\ 
3/2$^{+}$ & 1.919 & 931.56 & 5/2$^{+}$ & 1.918 & 922.49 & 3/2$^{+}$ & 1.922
& 927.99 \\ 
5/2$^{+}$ & 1.919 & 925.93 & 3/2$^{+}$ & 1.919 & 926.97 & 5/2$^{+}$ & 1.922
& 927.99 \\ 
1/2$^{+}$ & 2.054 & 983.77 & 1/2$^{+}$ & 2.069 & 964.19 & 1/2$^{+}$ & 2.110
& 1020.20 \\ 
3/2$^{-}$ & 2.379 & 1886.43 & 3/2$^{-}$ & 2.368 & 1822.55 & 3/2$^{-}$ & 2.422
& 2000.70 \\ 
5/2$^{-}$ & 2.719 & 1479.40 & 5/2$^{+}$ & 2.719 & 1476.00 & 5/2$^{-}$ & 2.271
& 1472.40 \\ 
5/2$^{+}$ & 2.960 & 989.56 & 3/2$^{+}$ & 2.998 & 1033.10 &  &  &  \\ 
\end{tabular}
\end{ruledtabular}
\end{table}%

\begin{figure}[ht]
\begin{center}
\includegraphics[width=0.75\textwidth]{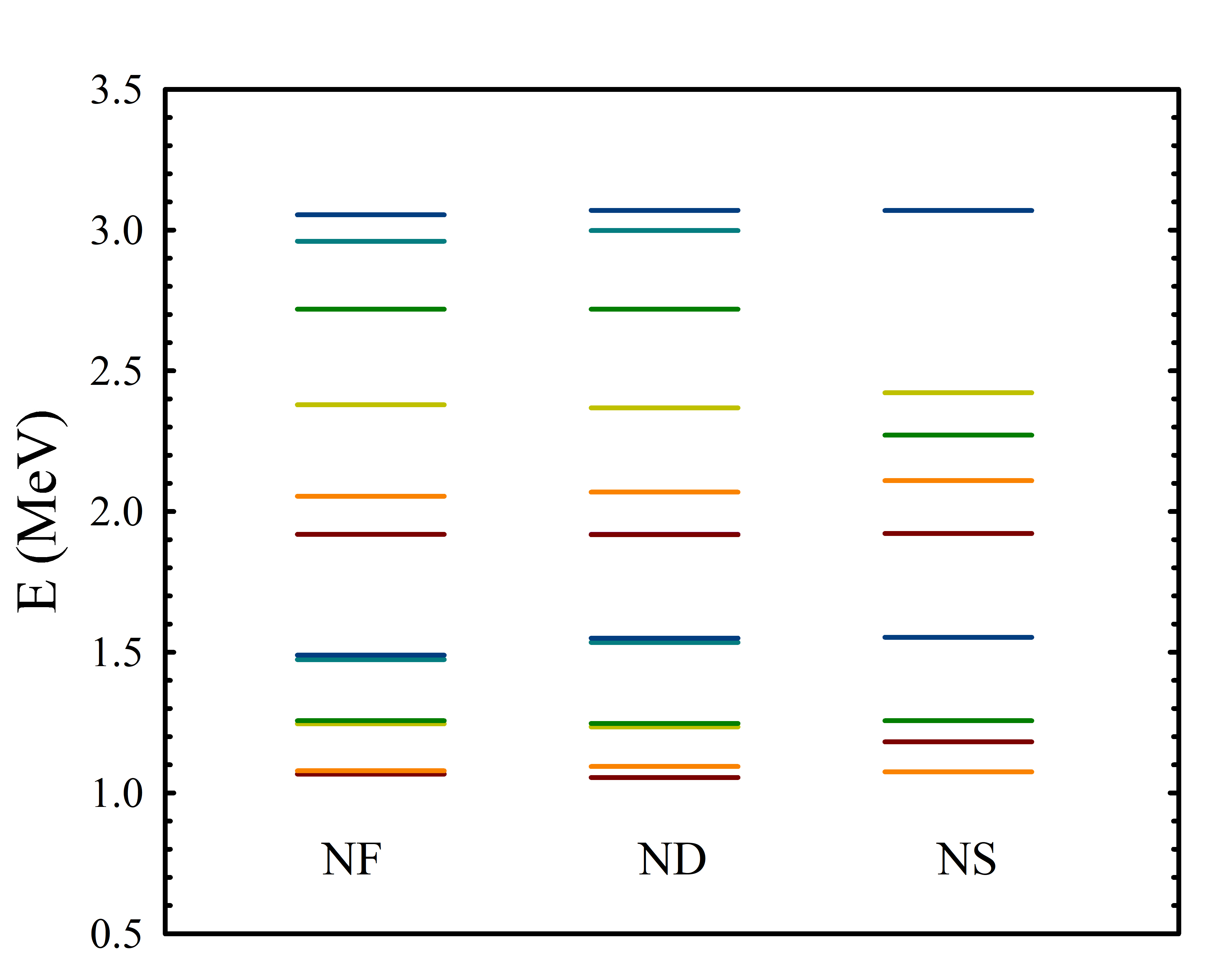}%
\caption{\label{Fig:Figure13}
Spectrum of resonance states in $_{\Lambda}^{7}$Be determined with
the YNG-NF, YNG-ND and YNG-NS potentials.}%
\end{center}
\end{figure}

The relative position of the resonance states in the three-cluster continuum of
$_{\Lambda}^{7}$Be is shown in Fig. \ref{Fig:Figure13}. Comparing the
spectrum of $_{\Lambda}^{7}$Be with one of $_{\Lambda}^{7}$He (Fig. \ref{Fig:Figure10}) we can see that the Coulomb interaction shifted up all resonance states. We also see that the shell-like structure of resonance states in $_{\Lambda}^{7}$Be is not as prominent as in $_{\Lambda}^{7}$He.%

\subsection{Effects of the Coulomb interaction}

Now we consider in more detail the effects of the Coulomb interaction on the bound and resonance states
of $_{\Lambda}^{7}$He and $_{\Lambda}^{7}$Be. As was pointed out, we
selected input parameters in such a way that potential energies of these
nuclei, generated by nucleon-nucleon and nucleon-hyperon interactions, are
equal. The difference in position of the bound and resonance states is totally
determined by the Coulomb interaction only.

\begin{figure}[ht]
\begin{center}
\includegraphics[width=0.75\textwidth]{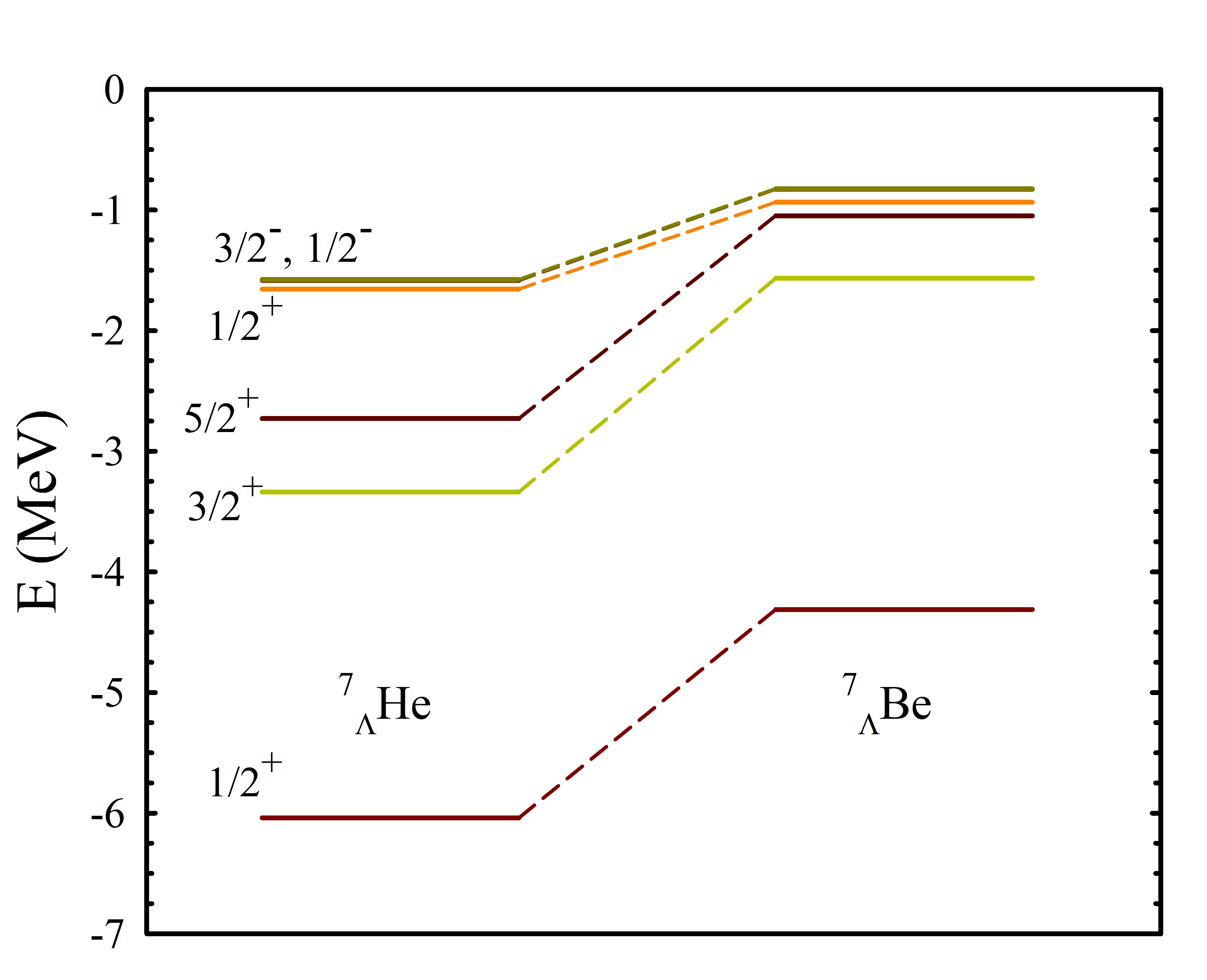}%
\caption{\label{Fig:Figure14}
Spectra of bound states of $_{\Lambda}^{7}$He and $_{\Lambda}^{7}$Be
obtained with the YNG-NF potential.}%
\end{center}
\end{figure}

We start with the analysis of the influence of the Coulomb interaction on the bound
states. In Fig. \ref{Fig:Figure14} we show the relative position of the
bound states in $_{\Lambda}^{7}$He and $_{\Lambda}^{7}$Be. They were obtained
with the YNG-NF potential. One can see that the Coulomb interaction has
strong effects on the 1/2$^{+}$, 3/2$^{+}$, 5/2$^{+}$ bound states, as it
significantly shifts the energies of bound states (approximately by 1.7 MeV) in
$_{\Lambda}^{7}$Be with respect to the position in $_{\Lambda}^{7}$He. The
Coulomb interaction has moderate effects on the second 1/2$^{+}$ bound state
and on the 1/2$^{-}$ and 3/2$^{-}$ bound states. The shift of the energy of these
states is approximately 0.7 MeV. The results, presented in Tables
\ref{Tab:Spectr7HHe3PotO} and \ref{Tab:Spectr7HBe3PotO}, indicate that the
strong effects of the Coulomb interaction are observed for the deeply bound compact bound states (with
the mass rms radii $R_{m}<$
2.9 fm) and moderate effects are observed for the weakly bound dispersed
(non-compact) bound states (their mass rms radii $R_{m}$ exceed
4.4 fm). The dispersed states have a small bound energy and lie very close to
the three-cluster decay threshold.%

\begin{figure}[ht]
\begin{center}
\includegraphics[width=0.75\textwidth]{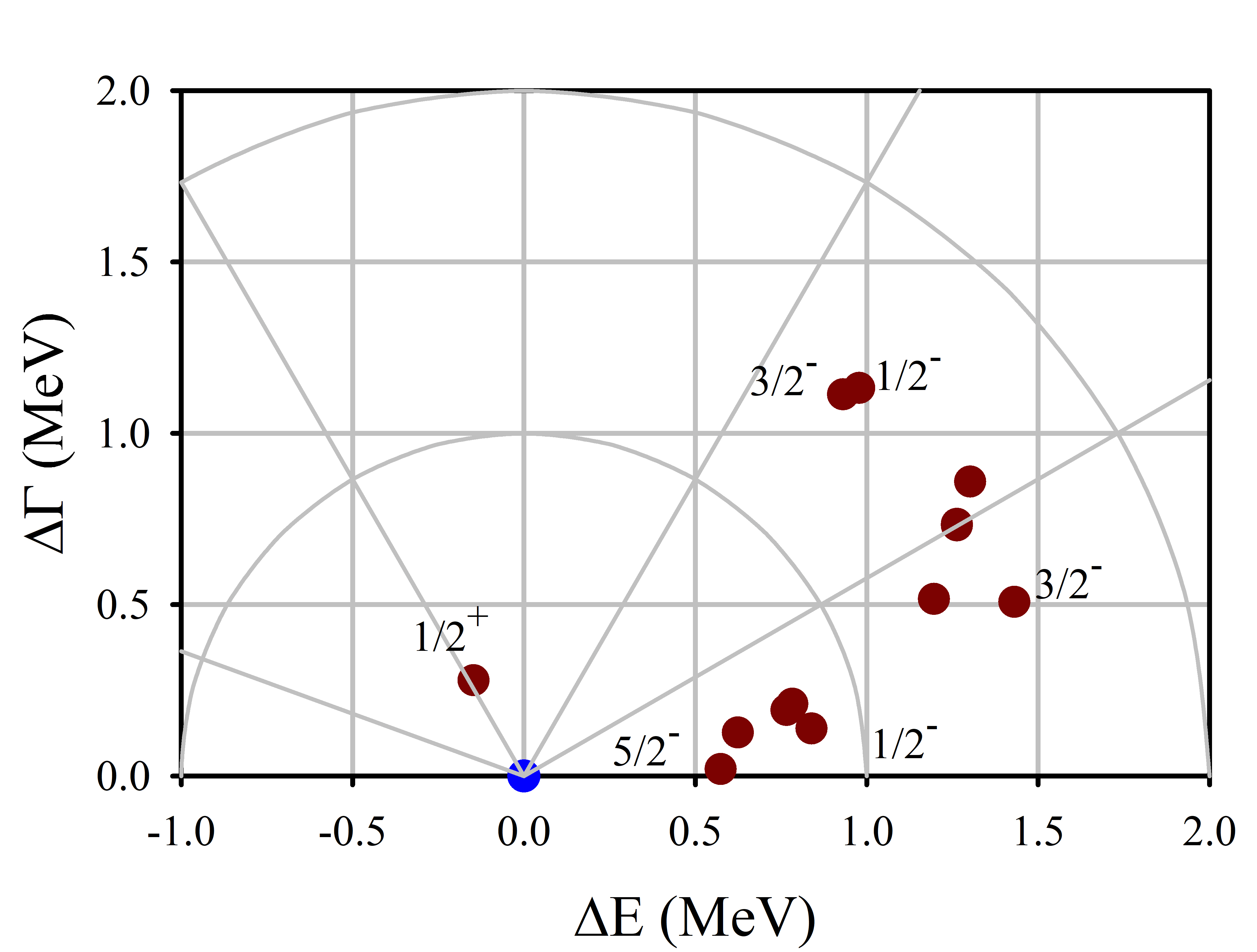}%
\caption{\label{Fig:Figure15}
Effects of the Coulomb interaction on the parameters of resonance states
in $_{\Lambda}^{7}$He and $_{\Lambda}^{7}$Be.}%
\end{center}
\end{figure}

Fig. \ref{Fig:Figure15} presents the visualization of effects of
the Coulomb interaction in hypernuclei $_{\Lambda}^{7}$He and $_{\Lambda
}^{7}$Be. For this visualization, we used an algorithm suggested in 
\cite{2020NuPhA.99621692D} to analyze the shift of the energies and widths of
resonance states in two- and three-cluster continua of mirror nuclei. For
resonance states having the same total angular momentum $J$ and parity $\pi$,
we determine the shift of the energy $\Delta E=E\left(  _{\Lambda}%
^{7}\text{Be}\right)  -E\left(  _{\Lambda}^{7}\text{He}\right)  $ and the
shift of the total width $\Delta\Gamma=\Gamma\left(  _{\Lambda}^{7}%
\text{Be}\right)  -\Gamma\left(  _{\Lambda}^{7}\text{He}\right)  $ of the
resonance states caused by the Coulomb interaction. In Fig. \ref{Fig:Figure15}, the red circles  indicate
the position of the $_{\Lambda}^{7}$Be resonance states on a two-dimensional plain
with respect to the position of the corresponding resonance states in $_{\Lambda}^{7}%
$He, indicated by the blue circles.

Four possible scenarios for the motion of resonance states caused by the Coulomb
interaction have been suggested in \cite{2020NuPhA.99621692D}. The first
scenario is for resonance states with larger energy and large widths (the
first quadrant of the plot), the second scenario unites  resonance states which have
smaller energy and larger width (the second quadrant), when the Coulomb
interaction decreases both energy and width, they belong to the third scenario
(the third quadrant), and the last fourth scenario is for resonance states with larger
energy but smaller width (the fourth quadrant). As we can see in Fig.
\ref{Fig:Figure15}, the Coulomb interaction in hypernuclei
 $_{\Lambda}^{7}$He and $_{\Lambda}^{7}$Be realizes the first and  second
scenarios, most of the resonance states determined realize the first scenario, and only
one resonance state (1/2$^{+}$) exhibits the second scenario. All resonance
states, which lie inside the circle with a radius of one MeV, we consider as
resonances with small effects of the Coulomb interaction, while resonance
outside this circle we consider as resonance states which undergo the strong
effects of the Coulomb interaction. It is interesting to note that resonance
states with small effects of the Coulomb interaction experience very small
shifts of the total width ($\Delta\Gamma<$0.3 MeV).

\subsection{Partial widths}

Now we proceed to study the partial widths of resonance states. They possess very
important information on resonance states and, in particular, they allow us to
determine dominant channels of their decay. We concentrated our attention on
most narrow resonance states.

In Table \ref{Tab:PartWidths7HHe12M52M} we show the partial widths of the narrow
1/2$^{-}$ and 5/2$^{-}$ resonance states in $_{\Lambda}^{7}$He, determined
with the NF potential. The dominant decay channel of the 1/2$^{-}$ resonance
state exhausts almost 97\% of all possible ways of decay. The lambda hyperon
escapes this resonance state with mainly zero angular momentum, and the relative
angular momentum of the two-cluster subsystem $^{4}$He+$^{2}$n is equal $l_{x}=$1.
Table \ref{Tab:PartWidths7HHe12M52M} shows that the 1/2$^{-}$ resonance state
is formed in the state with the total orbital momentum $L$=1 and the total spin
$S$=1/2. The  5/2$^{-}$ resonance state is formed by the total orbital
momentum $L$=3, channels with the hypermomentum $K$=3 describe the main ways of
its decay. The lambda hyperon escapes this resonance state primarily with the
orbital momentum  $l_{y}=$3. %

\begin{table}[b]
\caption{\label{Tab:PartWidths7HHe12M52M}%
Partial widths of decay of the narrow 1/2$^-$ ($E$=0.061 MeV) and 5/2$^-$ ($E$=0.493 MeV)  resonance states in $^7_{\Lambda}$He.}%
\begin{ruledtabular}
\begin{tabular}{cccccccc}
& \multicolumn{7}{c}{$E$=0.061 MeV, $\ \Gamma$=2.146 keV}\\
$i$ & $\Gamma_{i}$, keV & $\Gamma_{i}/\Gamma$, \% & $K$ & $l_{y}$ & $l_{x}$ &
$L$ & $S$\\\hline
1 & 2.081 & 96.94 & 1 & 0 & 1 & 1 & 1/2\\
2 & 0.065 & 3.05 & 1 & 1 & 0 & 1 & 1/2\\
3 & 0.0001 & 0.01 & 3 & 0 & 1 & 1 & 1/2\\\hline
& \multicolumn{7}{c}{$E$= 0.493 MeV, $\ \Gamma$=5.094 keV}\\\hline
1 & 4.035 & 79.22 & 3 & 3 & 0 & 3 & 1/2\\
2 & 0.938 & 18.42 & 3 & 1 & 1 & 3 & 1/2\\
3 & 0.089 & 1.76 & 3 & 1 & 2 & 3 & 1/2\\
4 & 0.022 & 0.43 & 3 & 0 & 3 & 3 & 1/2\\
5 & 0.003 & 0.07 & 5 & 3 & 0 & 3 & 1/2\\
\end{tabular}
\end{ruledtabular}
\end{table}%

In Table \ref{Tab:PartWidths7HBe32M52M} we show  the partial widths of the
narrow 5/2$^{-}$ and 3/2$^{-}$ resonance states in $_{\Lambda}^{7}$Be,
determined with the NF potential. Two channels with similar weights dominate
in both resonance states. The decay structure of the 5/2$^{-}$ resonance
states is similar to the structure of the 5/2$^{-}$ resonance state in
$_{\Lambda}^{7}$He, shown in Table \ref{Tab:PartWidths7HBe32M52M}. Indeed, the
total orbital momentum $L$=3  and the total spin $S$=1/2 dominate in both
resonance states. And the partial orbital momenta $l_{y}$=3 and  $l_{x}$=0
determine the main way of decay of both resonances. Two sets of 
partial orbital momenta $l_{y}$=0, $l_{x}$=1 and  $l_{y}$=1, $l_{x}$=0
determine the main way of decaying of the 3/2$^{-}$ resonance states in
$_{\Lambda}^{7}$Be.%

\begin{table}[b]
\caption{\label{Tab:PartWidths7HBe32M52M}%
Partial widths of decay of the 5/2$^-$ and 3/2$^-$ resonance states in
$^7_{\Lambda}$Be.}%
\begin{ruledtabular}
\begin{tabular}{cccccccc}
& \multicolumn{7}{c}{$E$=1.067 MeV, $\ \Gamma$=25.187 keV}\\
$i$ & $\Gamma_{i}$, keV & $\Gamma_{i}/\Gamma$, \% & $K$ & $l_{y}$ & $l_{x}$ &
$L$ & $S$\\\hline
1 & 21.179 & 84.09 & 3 & 3 & 0 & 3 & 1/2\\
2 & 3.645 & 14.47 & 3 & 2 & 1 & 3 & 1/2\\
3 & 0.213 & 0.84 & 3 & 1 & 2 & 3 & 1/2\\
4 & 0.060 & 0.24 & 3 & 3 & 0 & 3 & 1/2\\
5 & 0.045 & 0.18 & 3 & 4 & 1 & 3 & 1/2\\\hline
& \multicolumn{7}{c}{$E$= 1.246 MeV, $\ \Gamma$= 477.267 keV}\\\hline
1 & 404.603 & 84.78 & 1 & 0 & 1 & 1 & 1/2\\
2 & 66.237 & 13.88 & 1 & 1 & 0 & 1 & 1/2\\
3 & 4.314 & 0.90 & 3 & 0 & 1 & 1 & 1/2\\
4 & 1.874 & 0.39 & 3 & 1 & 0 & 1 & 1/2\\
5 & 0.160 & 0.03 & 5 & 1 & 2 & 1 & 1/2\\
\end{tabular}
\end{ruledtabular}
\end{table}%

In Table \ref{Tab:PartWidths7HLi12P} we show the partial widths of the two
narrowest 1/2$^{+}$ resonance states in $_{\Lambda}^{7}$Li, determined with
the NF potential. One can see that two channels contribute greatly to the decay of the
first resonance state (with weights 89.6\% and 9.8\%) and one channel significantly contributes to the decay of the second
1/2$^{+}$ resonance state (with weight 98.9\%). The total orbital momentum $L$=2 and the total spin
$S$=3/2 dominate in both resonance states, while the $LS$ combinations
(0,1/2), (1,1/2), (1,3/2) give a very small contribution. Channels with
a total hypermomentum $K$=2 give the largest contributions to the total
width of the 1/2$^{+}$ resonance states. The lambda hyperon escapes the first
resonance state mainly with the orbital momentum $l_{y}$=2 and prefers to
escape the second resonance state with $l_{y}$=0.%

\begin{table}[b]
\caption{\label{Tab:PartWidths7HLi12P}%
Partial widths of decay of the 1/2$^+$ ($E$=0.333 MeV)  and 1/2$^+$ ($E$=0.544 MeV)  resonance states in
$^7_{\Lambda}$Li.}%
\begin{ruledtabular}
\begin{tabular}{cccccccc}
& \multicolumn{7}{c}{$E$= 0.333 MeV, $\ \Gamma$= 1.54\ keV}\\
$i$ & $\Gamma_{i}$, keV & $\Gamma_{i}/\Gamma$, \% & $K$ & $l_{y}$ & $l_{x}$ &
$L$ & $S$\\\hline
1 & 1.380 & 89.58 & 2 & 2 & 0 & 2 & 3/2\\
2 & 0.150 & 9.75 & 2 & 1 & 1 & 2 & 3/2\\
3 & 0.008 & 0.53 & 2 & 0 & 2 & 2 & 3/2\\
4 & 0.0009 & 0.06 & 4 & 2 & 0 & 2 & 3/2\\\hline
& \multicolumn{7}{c}{$E$= 0.544 MeV, $\ \Gamma$= 11.12 keV}\\\hline
1 & 10.996 & 98.91 & 2 & 0 & 2 & 2 & 3/2\\
2 & 0.097 & 0.87 & 2 & 2 & 0 & 2 & 3/2\\
3 & 0.019 & 0.17 & 4 & 0 & 2 & 2 & 3/2\\
4 & 0.004 & 0.04 & 2 & 1 & 1 & 2 & 3/2\\
\end{tabular}
\end{ruledtabular}
\end{table}%

\subsection{Nature of resonance states}

Partial widths along with correlation function and weights of oscillator
shells in the resonance wave functions help us to understand the nature of the found resonances. To this list of quantities related to excitation and
decay of three-cluster resonance states, we can add the average distances
between clusters. They are calculated in the same fashion as for bound states.
In Table \ref{Tab:ResonsHA7AvDist} we show the average distance between clusters for the
narrow resonance states in $_{\Lambda}^{7}$He, $_{\Lambda}^{7}$Li and
$_{\Lambda}^{7}$Be, which were determined with the NF potential. The 1/2$^-$ and
3/2$^-$ resonance states in $_{\Lambda}^{7}$He, having approximately the same
energy and width, also have approximately the same average distances between
clusters. The base of the three-cluster triangles in these states is small, 
slightly more than 5 fm, while the height of the triangles is very large
($\approx$15 fm). Thus, these triangles are acute. This is partially
stipulated by the fact 
that both resonance states lie very close to the three-cluster
threshold and their wave functions are suppressed at small distances. It is
interesting that the triangles associated with the narrow 5/2$^{-}$ resonance
states in $_{\Lambda}^{7}$He and $_{\Lambda}^{7}$Be have a large base but a
small height. These triangles are obtuse. It is interesting to compare the
binary average distances $R_{B}$, presented in Table \ref{Tab:ResonsHA7AvDist} with the analogous values
determined in a two-cluster model. Using an algorithm described in 
\cite{2023UkrJPh..68..3K}, we determined the average distances between the
clusters $^{4}$He and $d$ both for the ground state of $^{6}$Li and for the
narrow 3$^{+}$ resonance state. With the HNP potential, they are 4.88  and
6.73 fm, respectively. For all narrow resonance states in $_{\Lambda}^{7}$Li,
the average distances between the clusters $R_{B}$ vary from 8.17 to 9.23 fm, which is
larger than the average distance of $^{6}$Li the ground state and in the  3$^{+}$ narrow
resonance state. If we compare the average distance between the clusters $^{4}$He
and $d$ in the 1$^{+}$ ground state of $^{6}$Li, determined in a two-cluster
approximation, with the average distances $R_{B}$ in the deeply bound states
of $_{\Lambda}^{7}$Li, we see that the presence of  the lambda hyperon
substantially compresses $^{6}$Li, since its radius $R_{B}$ varies from 2.88 to 3.33 fm,
which is much smaller than 4.88 fm. This phenomenon has been discussed in
Ref. \cite{2001PhRvL..86.1982T}, where experimental evidence for the shrinkage of
$^{6}$Li in the ground state of $_{\Lambda}^{7}$Li has been experimentally
detected.%

\begin{table}[b]
\caption{\label{Tab:ResonsHA7AvDist}%
The size and shape parameters of the narrowest resonance states in
$^7_{\Lambda}$He, $^7_{\Lambda}$Li and $^7_{\Lambda}$Be.}%
\begin{ruledtabular}
\begin{tabular}{cccccc}
$_{\Lambda}^{A}Z$ & $J^{\pi}$ & $E$, MeV & $\Gamma$, keV & $R_{\Lambda}$, fm &
$R_{B}$, fm\\\hline
$_{\Lambda}^{7}$He & 1/2$^{-}$ & 0.061 & 2.16 & 14.99 & 5.39\\
& 3/2$^{-}$ & 0.043 & 1.09 & 14.72 & 5.42\\
& 5/2$^{-}$ & 0.493 & 5.09 & 5.22 & 10.30\\
$_{\Lambda}^{7}$Be & 5/2$^{-}$ & 1.067 & 25.19 & 5.27 & 10.51\\
$_{\Lambda}^{7}$Li & 1/2$^{+}$ & 0.333 & 1.54 & 4.98 & 8.17\\
& 5/2$^{+}$ & 0.519 & 8.35 & 6.71 & 9.09\\
& 3/2$^{+}$ & 0.520 & 8.43 & 6.76 & 9.13\\
& 1/2$^{+}$ & 0.544 & 11.12 & 11.09 & 7.39
\end{tabular}
\end{ruledtabular}
\end{table}%

\begin{figure}[ht]
\begin{center}
\includegraphics[width=0.75\textwidth]{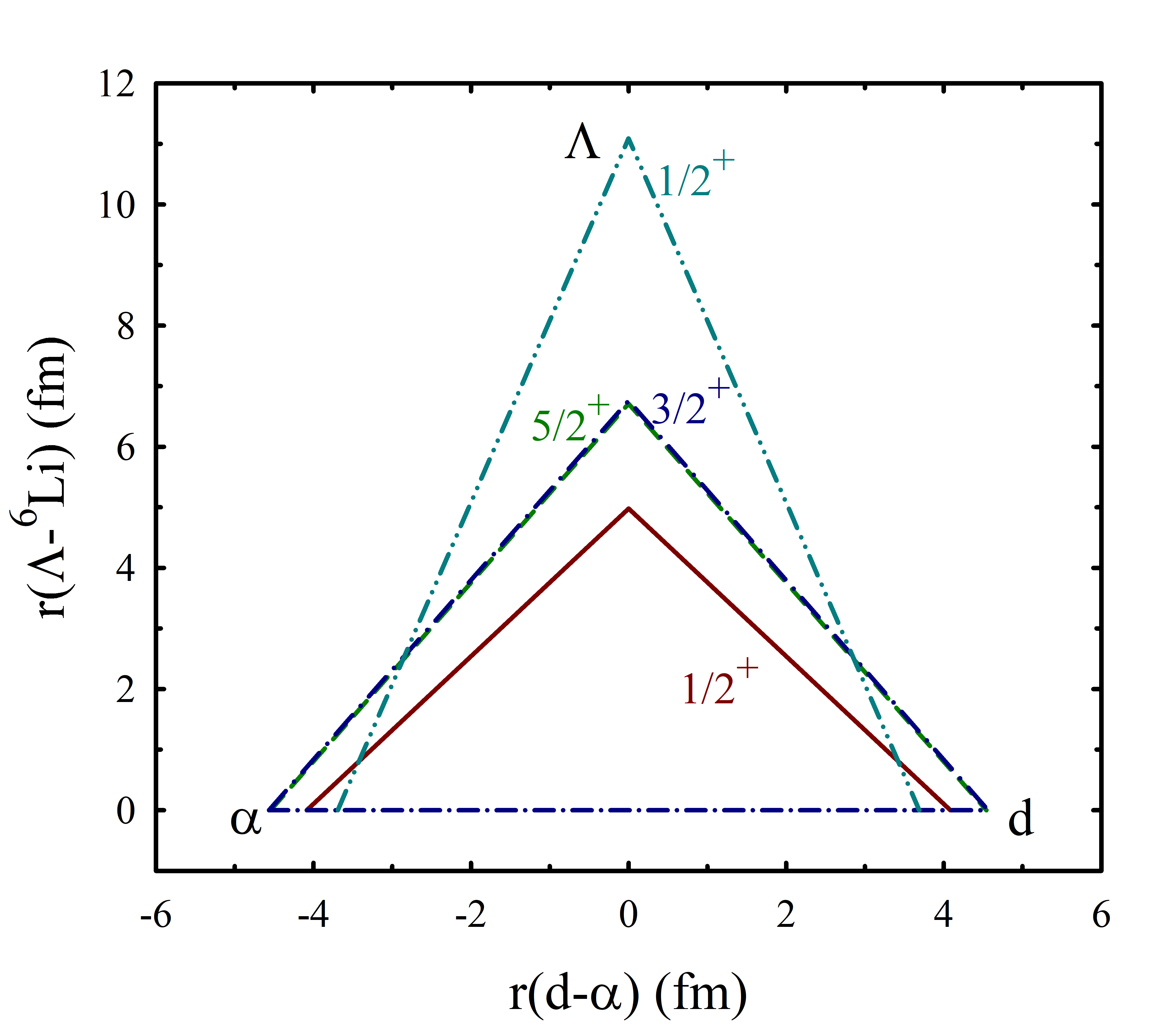}%
\caption{\label{Fig:Figure16}
Effective size and shape of triangles for narrowest resonance states in
\ $_{\Lambda}^{7}$Li.}%
\end{center}
\end{figure}

In Fig. \ref{Fig:Figure16} we visualize the shape and size of
triangles, connecting three clusters $^{4}$He, d and $\Lambda$, for resonance
states in $_{\Lambda}^{7}$Li. In the first 1/2$^{+}$ resonance state, the lambda
hyperon is relatively close to the center of mass of the two-cluster $^{6}$Li
subsystem, in the second 1/2$^{+}$ resonance state the lambda hyperon is far away
from $^{6}$Li, however, in this state the distance between the alpha particle and the
deuteron is smallest among the four narrowest resonance states, detected in
$_{\Lambda}^{7}$Li.%

\begin{figure}[ht]
\begin{center}
\includegraphics[width=0.75\textwidth]{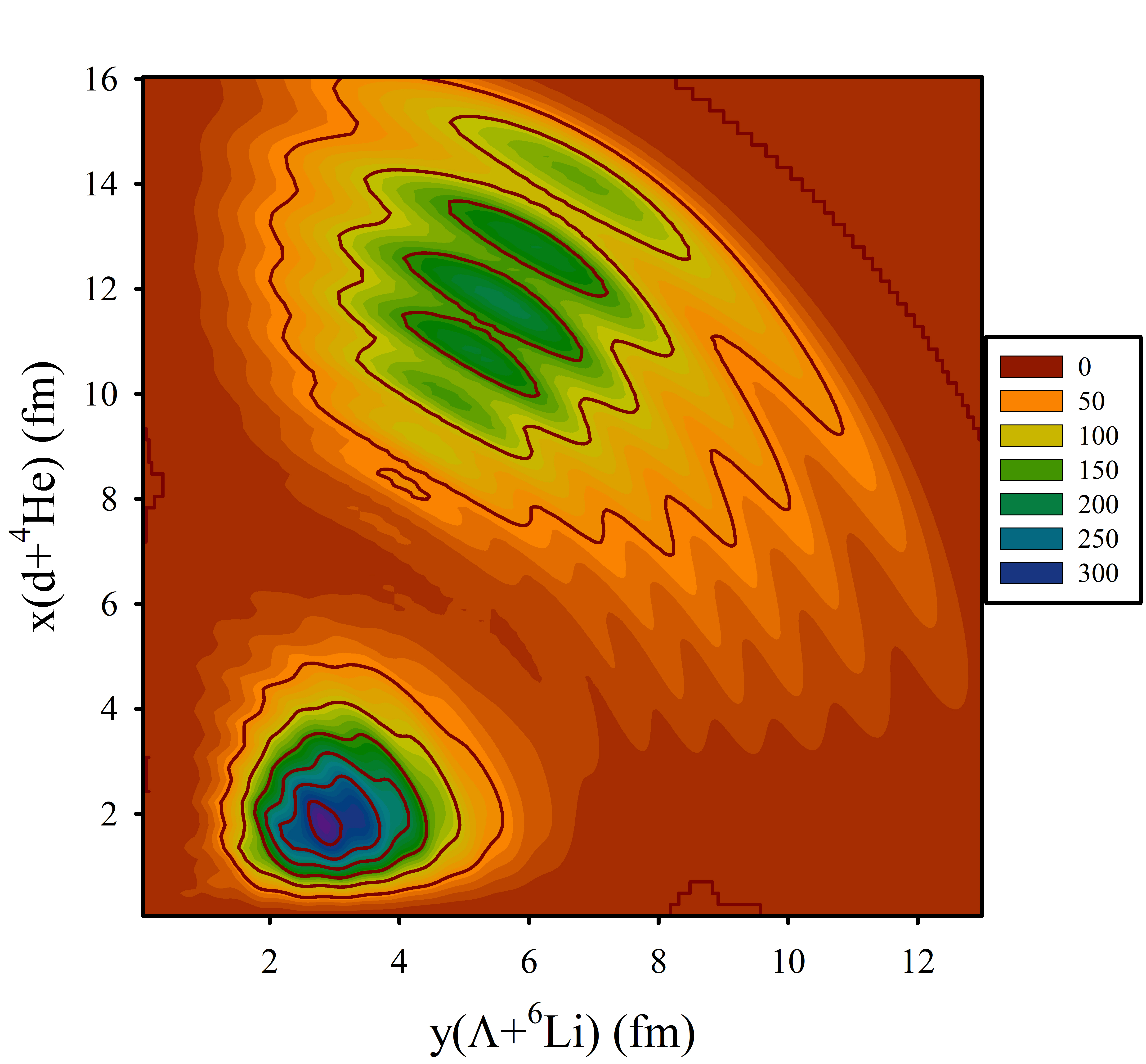}%
\caption{\label{Fig:Figure17}
Correlation function of the 1/2$^{+}$ resonance state in $_{\Lambda
}^{7}$Li, obtained with the YNG-NF potential.}%
\end{center}
\end{figure}

\begin{figure}[ht]
\begin{center}
\includegraphics[width=0.75\textwidth]{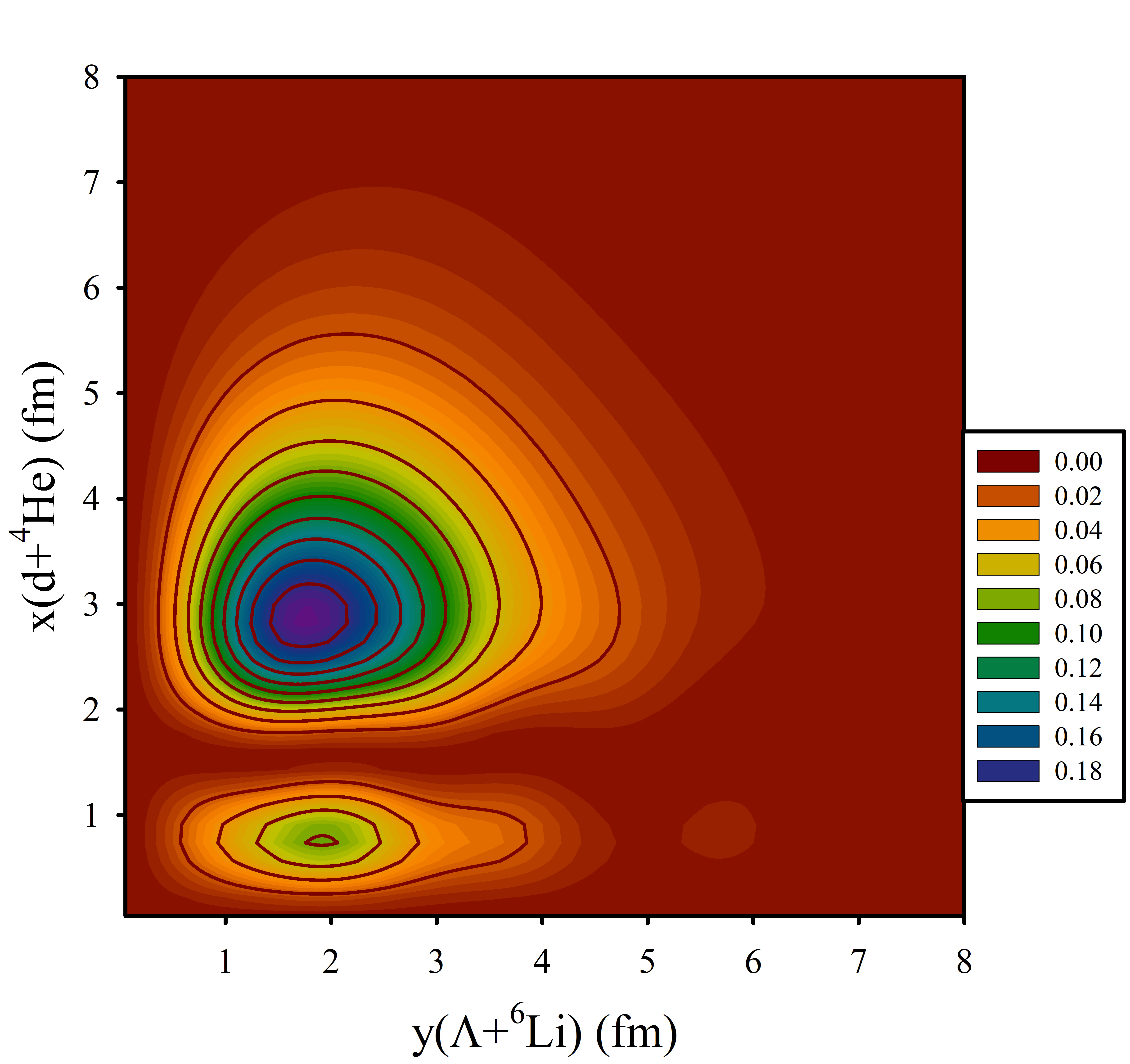}%
\caption{\label{Fig:Figure18}%
Correlation function for the 1/2$^{+}$ ground state of $_{\Lambda
}^{7}$Li, constructed with the YNG-NF potential.}%
\end{center}
\end{figure}

In Fig. \ref{Fig:Figure17} we show the correlation function for the
narrowest resonance 1/2$^{+}$ state in $_{\Lambda}^{7}$Li which have the
parameters $E$=0.333 MeV and $\Gamma$= 1.54 keV. This figure indicates that
the most probable distance between an alpha particle and a deuteron is
approximately 1.72 fm and the displacement of the lambda particle from the center of mass
of $^{6}$Li is approximately 2.85 fm. This is a very compact configuration.
It is interesting to compare the correlation functions for the narrowest
resonance state with the configuration function of the ground 1/2$^{+}$ state,
which is shown in Fig \ref{Fig:Figure18}. The ground state is
also a fairly compact state with the reverse position of clusters, where the
most probable distance between an alpha particle and a deuteron is 2.89 fm and the
distance between the lambda hyperon and the center of mass of $^{6}$Li is
approximately 1.75 fm.%

\begin{figure}[ht]
\begin{center}
\includegraphics[width=0.75\textwidth]{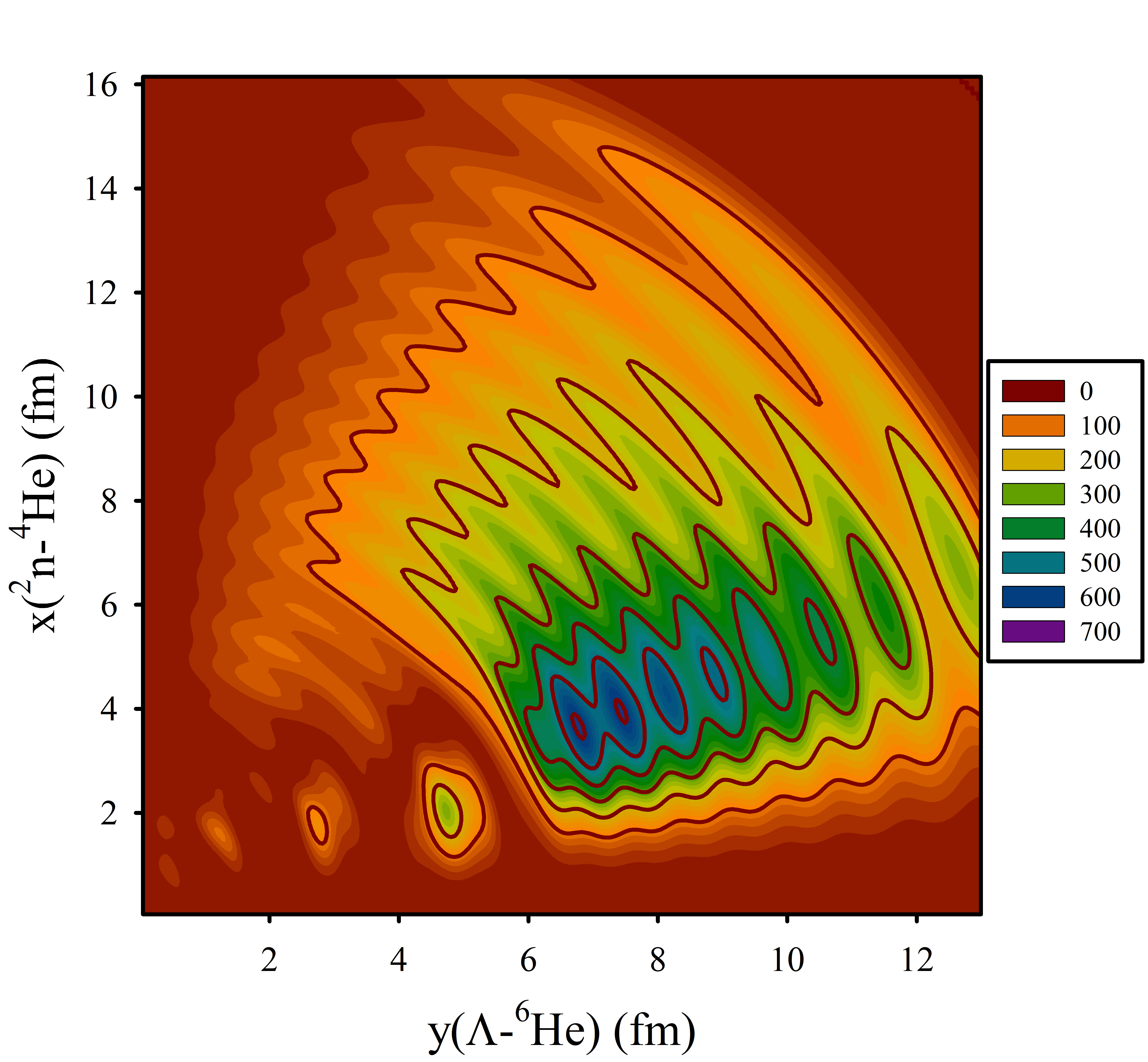}%
\caption{\label{Fig:Figure19}%
Correlation function of the 3/2$^{-}$ narrowest resonance state in
$_{\Lambda}^{7}$He, determined with the YNG-NF potential.}%
\end{center}
\end{figure}

The correlation function for the 3/2$^{-}$ narrowest resonance state in
$_{\Lambda}^{7}$He is shown in Fig. \ref{Fig:Figure19}.  The width of
this resonance state is 1.1 keV and slightly less than the width of the
1/2$^{+}$ resonance state in  $_{\Lambda}^{7}$Li, whose correlation function is shown in Fig.
\ref{Fig:Figure17}.  Fig. \ref{Fig:Figure19} indicates that the
small distances between the clusters $^{2}n$ and $^{4}$He are dominant in the
wave function of the 3/2$^{-}$ resonance state, and it also reveals that the
lambda hyperon resides mainly at large distances from $^{6}$He. The
correlation function  also demonstrates that the 3/2$^{-}$ resonance state
decays by emitting the lambda hyperon, which moves far from a two-cluster
subsystem $^{2}$n+$^{4}$He, and these clusters move apart with a
small relative velocity.%

\begin{figure}[ht]
\begin{center}
\includegraphics[width=0.75\textwidth]{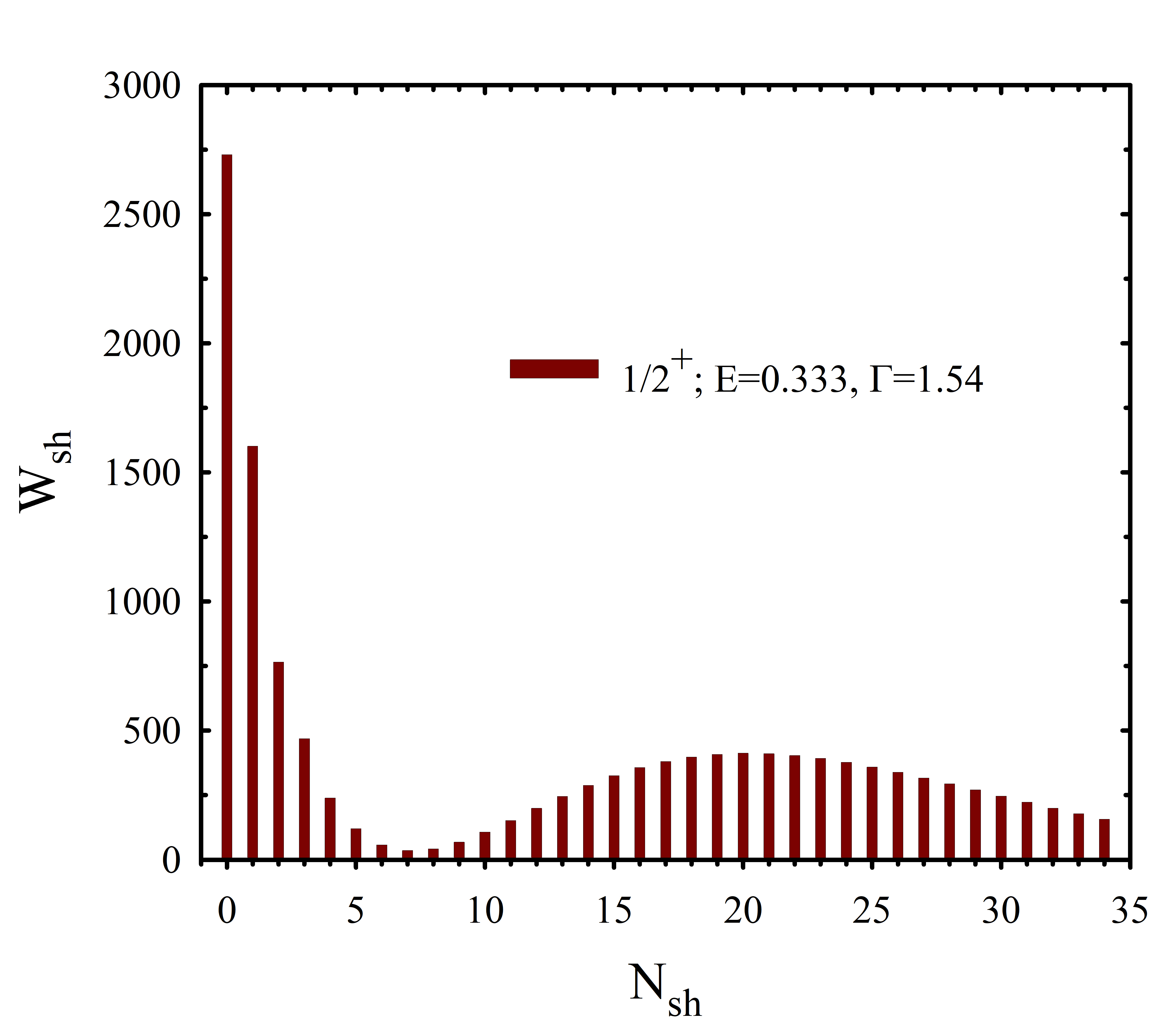}
\caption{\label{Fig:Figure20}
Weights $W_{sh}$ of different oscillator shells $N_{sh}$ in the wave
function of the narrowest 1/2$^{+}$ resonance state in $_{\Lambda}^{7}$Li.}%
\end{center}
\end{figure}

Another view of the structure of the narrow resonance state is proposed by the
weights $W_{sh}$ of different oscillator shells $N_{sh}$ in the wave function of a
resonance state. Fig. \ref{Fig:Figure20} displays such
quantities for the narrowest 1/2$^{+}$ resonance state in $_{\Lambda}^{7}$Li.
One can see that these weights have very large amplitudes and 
are represented by oscillator shells with a small value of $N_{sh}$.
Comparing Fig. \ref{Fig:Figure20} with figures in Ref.
\cite{2018PhRvC..98b4325V}, where weights of oscillator shells have been
demonstrated for the Hoyle and Hoyle-analog states in light nuclei, we came to the
conclusion that the 1/2$^{+}$ resonance state is quit comparable with the
structure of the narrow resonance states in three-cluster continuum and also
with the Hoyle-analog states, which have a total width of the order of a few eV.%

\begin{figure}[b]
\begin{center}
\includegraphics[width=0.75\textwidth]{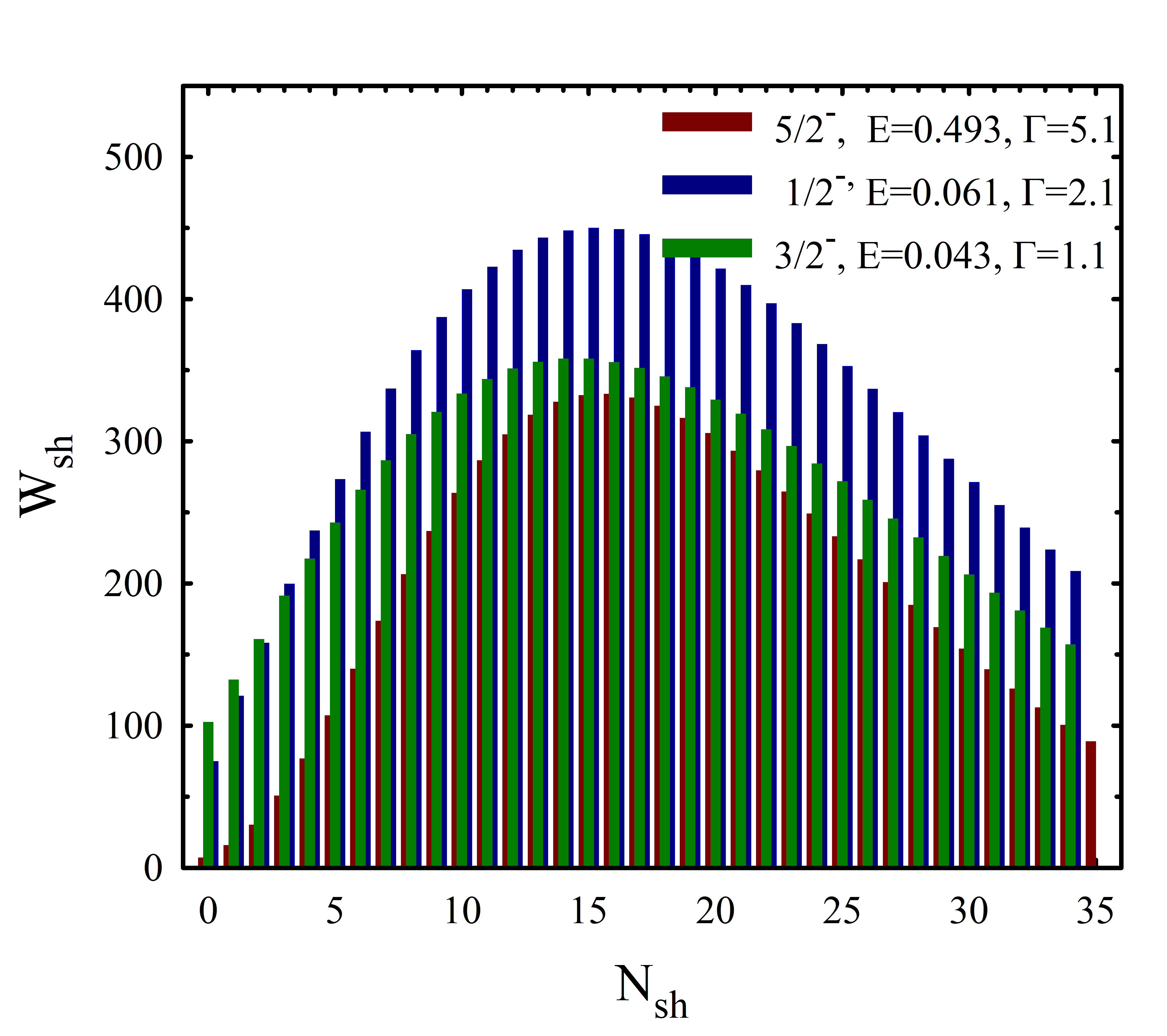}%
\caption{\label{Fig:Figure21}
Weights of different oscillator shells in wave functions of the
narrowest resonance states in $_{\Lambda}^{7}$He. Results are obtained with the YNG-NF potential.}%
\end{center}
\end{figure}

The structure of wave functions of the narrowest resonance states in $^{7}$He is
shown in Fig. \ref{Fig:Figure21}, where the weights $W_{sh}$ of
oscillator shells $N_{sh}$ are displayed for the 1/2$^{-}$, 3/2$^{-}$ and
5/2$^{-}$ resonances. The weights of the oscillator shells in these wave
functions reflect a similarity in their behavior. First, small values of
$W_{sh}$ for oscillator shells with small $N_{sh}$, second, large
amplitudes $W_{sh}>250$ for relatively large shells 10$<N_{sh}<$25.

Comparing shell weights for the 1/2$^{+}$ resonance state in $_{\Lambda}^{7}
$Li (Fig. \ref{Fig:Figure20}) and shell weights for the
1/2$^{-}$, 3/2$^{-}$ and 5/2$^{-}$ resonance states in $_{\Lambda}^{7}$He
(Fig. \ref{Fig:Figure21}), one may conclude that something is wrong with these results, as 
the narrow resonance states with close values of the
total width have quite different behavior of their wave functions. To resolve
this doubt, we need to pay attention that these resonance states lie very
close to the three-cluster threshold. One needs to recall how two- and
three-particle wave functions of low-energy states behave at small
distances. In this case, the product $\left(  kr\right)  $  is small (here,  
$r$ is a distance between two particles, and $k=\sqrt{\frac{2mE}{\hbar^{2}}}$
is the wave number) and the wave function of the scattering state behaves as%
\[
\Psi_{E,L}\left(  r\right)  \approx\left(  kr\right)  ^{L}.
\]
For a three-particle system, the wave function of the low-energy scattering state behaves
as%
\[
\Psi_{E,K}\left(  \rho\right)  \approx\left(  k\rho\right)  ^{K+1/2}.
\]
One can see that, for low-energy scattering, the wave functions of two- and
three-particle systems are suppressed at small distances. The larger the
quantum numbers $L$ or $K$, the smaller is a wave function at small distances.
This behavior of the wave functions is reflected in the behavior of the expansion
coefficients (\ref{eq:N033}) and also in the weights of the oscillator shells
(\ref{eq:N035}). This means that  the expansion coefficients $C_{n_{\rho},c}$
are very small when $n_{\rho}$ is small, and therefore the weights $W_{sh}$ are
small when $N_{sh}$ is small. We observe such behavior of weights $W_{sh}$ for a
low-energy resonance state, where the nonzero value of the hypermomentum $K$ is dominant.

\subsection{Comparing with other models}

In Ref. \cite{2015PhRvC..91e4316H}, the resonance structure of $_{\Lambda}%
^{7}$He has been studied within a four-cluster model and energies and widths
of resonance states in two-, three-, and four-cluster continua were determined
by using the complex scaling method.  To our knowledge, this is the only 
publication in which the resonance states of  $_{\Lambda}^{7}$He, $_{\Lambda}^{7}$Li
and $_{\Lambda}^{7}$Be were detected. Two very broad resonance states were
found in $_{\Lambda}^{7}$He, they lie very close to the four-cluster threshold
$^{4}$He+$n$+$n$+$\Lambda$.  These resonance states were found in states
3/2$^{+}$ and 5/2$^{+}$ and their parameters are $E$=0.03 MeV,  $\Gamma$ =
1.03 MeV and $E$=0.07 MeV, $\Gamma$ = 1.01 MeV, respectively. In our model, the YNG-NF potential generates
these resonance states at an energy $E$=0.656 MeV above the
three-cluster threshold $^{4}$He+$^{2}n$+$\Lambda$, and their  total width
equal to $\Gamma$ = 196 keV. In our opinion, it is important that different theoretical methods predict the existence of resonance states in the three- and four-cluster continuum of hypernucleus $_{\Lambda}^{7}$He. The difference in parameters of the revealed resonance states is related to different nucleon-nucleon  and nucleon-hyperon potentials employed in these models and also to different methods of determining their position.

\section{Conclusions\label{Sec:Concl}}

We have investigated the bound and resonance states of hypernuclei $_{\Lambda}%
^{7}$He, $_{\Lambda}^{7}$Li and $_{\Lambda}^{7}$Be. These hypernuclei were
considered as three-cluster systems $^{4}$He+$^{2}n$+$\Lambda$, $^{4}$%
He+$d$+$\Lambda$, $^{4}$He+$^{2}p$+$\Lambda$, respectively. We apply a microscopic
three-cluster model which correctly treats the Pauli principle for
nucleons and takes into account proper boundary conditions for the decay of a
compound system into three clusters. The latter allowed us to scan the
three-cluster continuum of these hypernuclei and find resonance states. For
self-consistency of our calculations with the available experimental data, we
tuned parameters of the nucleon-nucleon interaction to reproduce the bound state
energy of two-cluster subsystems, comprised of ordinary clusters. We also
slightly tuned the parameters of the nucleon-hyperon interaction 
to
reproduce the ground state energy of a compound hypernucleus. In such a way, we
"synchronized" the interaction between three constituent clusters in each
hypernucleus and then determined the energies of bound states, energies and widths of resonance states. 
It was demonstrated that the spectrum of bound states, determined in our model, is in satisfactory agreement with the results of the alternative three-cluster model, which employs the same $\Lambda N$ potentials.
A set
of very narrow and fairly wide resonance states emerged in our calculations.
The narrowest resonance states were detected in $_{\Lambda}^{7}$He and
$_{\Lambda}^{7}$Li and they have a total width less than 10 keV, their energy
does not exceed 0.60 MeV above a three-cluster threshold.
We assume that the narrowest resonance states  in $_{\Lambda}%
^{7}$He and $_{\Lambda}^{7}$Li, predicted by our model, can be experimentally detected. 

We studied in detail the effects of the Coulomb interaction on the bound states and on the
parameters of the resonance states. The Coulomb interaction was found to have a strong and weak impact on the
 bound and resonance states. The strongest effects were
discovered for the deeply bound states and for the narrowest resonance states, which 
 are fairly compact states.

\begin{acknowledgments}
This research has been partially funded by the Science Committee of the Ministry
of Education and Science of the Republic of Kazakhstan (Grant No. AP22683187,
the project title "Structure of the light nuclei and hypernuclei in
multi-channel and multi-cluster models") and received partial support from the
Program of Fundamental Research of the Physics and Astronomy Department of the
National Academy of Sciences of Ukraine (Project No. 0122U000889).

\end{acknowledgments}


\bibliography{paper7HA}

\end{document}